\documentclass[revtex4]{emulateapj}

\def\kms{km~s$^{-1}$}

\def\feh{$[{\rm Fe/H}]$}
\def\teff{$T_{\rm eff}$}
\def\logg{$\log \, g$}
\def\stripe{Stripe~82}
\def\kms{km~s$^{-1}$}

\usepackage[colorlinks,urlcolor=blue,citecolor=blue,linkcolor=blue]{hyperref}
\usepackage{booktabs}
\slugcomment{DRAFT \today}
\shorttitle{Stellar Parameters from Photometric Light Curves}
\shortauthors{Miller et.\ al.}

\begin{document}

\title{A Machine Learning Method to Infer Fundamental Stellar Parameters 
from Photometric Light Curves}

\author{A.~A.~Miller\altaffilmark{1,2,3*}, 
J.~S.~Bloom\altaffilmark{4,5},
J.~W.~Richards\altaffilmark{4,6},
Y.~S.~Lee\altaffilmark{7}, \\
D.~L.~Starr\altaffilmark{4,6},
N.~R.~Butler\altaffilmark{8},
S.~Tokarz\altaffilmark{9},
N.~Smith\altaffilmark{10},
and 
J.~A.~Eisner\altaffilmark{10}
}

\altaffiltext{1}{Jet Propulsion Laboratory, California Institute of Technology, 4800 Oak Grove Drive, 
    MS 169-506, Pasadena, CA 91109, USA}
\altaffiltext{2}{California Institute of Technology, Pasadena, CA 91125, USA}
\altaffiltext{3}{Hubble Fellow}
\altaffiltext{4}{Department of Astronomy, University of California, 
    Berkeley, CA 94720-3411, USA}
\altaffiltext{5}{Physics Division, Lawrence Berkeley National Laboratory, 
    Berkeley, CA 94720, USA}
\altaffiltext{6}{wise.io, Berkeley, CA 94704, USA}
\altaffiltext{7}{Department of Astronomy and Space Science, 
    Chungnam National University, Daejeon 305-764,  Republic of Korea}
\altaffiltext{8}{School of Earth \& Space Exploration, Arizona State    
    University, Tempe, AZ 85281, USA}
\altaffiltext{9}{Smithsonian Astrophysical Observatory, Cambridge, 
    MA 02138, USA}
\altaffiltext{10}{Steward Observatory, University of Arizona, Tucson, 
    AZ 85721, USA}
\altaffiltext{*}{E-mail: {\tt amiller@astro.caltech.edu}}

\begin{abstract}
    A fundamental challenge for wide-field imaging surveys is obtaining follow-up spectroscopic observations: there are $> 10^9$ photometrically cataloged sources, yet modern spectroscopic surveys are limited to $\sim$few$\times 10^6$ targets. As we approach the Large Synoptic Survey Telescope (LSST) era, new algorithmic solutions are required to cope with the data deluge. Here we report the development of a machine-learning framework capable of inferring fundamental stellar parameters (\teff, \logg, and \feh) using photometric-brightness variations and color alone. A training set is constructed from a systematic spectroscopic survey of variables with Hectospec/MMT. In sum, the training set includes $\sim$9000 spectra, for which stellar parameters are measured using the SEGUE Stellar Parameters Pipeline (SSPP). We employed the random forest algorithm to perform a non-parametric regression that predicts \teff, \logg, and \feh\ from photometric time-domain observations. Our final, optimized model produces a cross-validated root-mean-square error (RMSE) of 165~K, 0.39~dex, and 0.33~dex for \teff, \logg, and \feh, respectively. Examining the subset of sources for which the SSPP measurements are most reliable, the RMSE reduces to 125~K, 0.37~dex, and 0.27~dex, respectively, comparable to what is achievable via  low-resolution spectroscopy. For variable stars this represents a $\approx$12$-20$\% improvement in RMSE relative to models trained with single-epoch photometric colors. As an application of our method, we estimate stellar parameters for $\sim$54,000 known variables. We argue that this method may convert photometric time-domain surveys into pseudo-spectrographic engines, enabling the construction of extremely detailed maps of the Milky Way, its structure, and history.  
\end{abstract}

\keywords{methods: data analysis -- methods: statistical -- stars: general -- stars: statistics -- stars: variables: general -- surveys}

\section{Introduction}\label{intro}

New time-domain surveys have begun exploring everything from nearby extrasolar planets to the most distant known stellar explosions, and a veritable zoo of time-variable astrophysical phenomena in the space between (e.g., \citealt{Borucki10,law09,Gehrels04}). The volume of data and sheer breath of inquiry of existing surveys will eventually be dwarfed by the Large Survey Synoptic Telescope (LSST; \citealt{Ivezic08}), which will track the brightness variations of $\sim$20 billion sources throughout the Universe. The rapidly increasing rate at which we acquire and process observations for these surveys requires sophisticated algorithms capable of discovering and classifying new sources as well as, or better than, human experts. 

Machine-learning methods provide a promising avenue for the necessary abstraction of the discovery and classification process.\footnote{For a primer on machine learning we refer the interested reader to \citet{Hastie09}.} The algorithms defining these methods are data driven, built to learn relationships between observables and parameters of interest without relying on parametric physical models.

The learning is achieved using objects with known properties (such as a variable star classification or a galaxy redshift), which is called the training set. Once a machine-learning model has been trained, it can be rapidly applied to new data providing predictions of the quantities of interest. As more data are obtained, and the quality and scope of the training set are improved, the machine can refine its knowledge and model of the dataset, providing ever more accurate predictions. Furthermore, unlike humans, machine-learning models can nearly instantaneously and automatically produce predictions about new data via a fully scalable process.

The application of these statistical machine-learning approaches to photometric light curves has enabled high-fidelity classifications of stellar variables (e.g., \citealt{debosscher07,Dubath11,Richards12a}). Spectroscopic observations are typically required, however, for the precise measurement of fundamental stellar properties, particularly metallicity (\feh) and surface gravity (\logg).

There is now a decades-long history of studies using photometric measurements, which are relatively cheap to obtain, to estimate stellar properties that are typically inferred from spectroscopic measurements, which are expensive. It is well established that photometric colors are particularly useful for  estimating \teff, capable of producing a typical scatter of $<$0.01 dex relative to spectroscopic  measurements, even in cases where only a single photometric color is available (e.g., \citealt{Ivezic08a}).  Photometric estimates of surface gravity and metallicity, on the other hand, have proven more  challenging (e.g., \citealt{Brown11}). 

It was first recognized by \citet{Schwarzschild55} that stellar atmospheres with enhanced metal content produce less flux in the blue portion of the optical. Many studies have leveraged this fact to photometrically estimate metallicity (or \feh) using broadband, blue photometric colors, typically including either the Johnson $U$- or SDSS $u$-band. Broadband colors are capable of producing a typical scatter of $\approx$0.20 dex when restricted to FG stars \citep{Ivezic08a}, and $\approx$0.30 dex when no color restrictions are adopted (see \citealt{Kerekes13}). 

The most precise photometric estimates of stellar properties are determined using narrow- and medium-band filters, which are designed to be sensitive to both  metal- and surface gravity-dependent spectral features. The most prominent technique uses the $uvby\beta$ Str{\"o}mgren filters (see \citealt{Stromgren66} for a review), which has been demonstrated to produce a scatter of $\approx$0.10 dex for \feh\ relative to spectroscopic observations for FG stars \cite{Nordstrom04}. The $uvby\beta$ filters do not, however, provide precise estimates for late-type (KM) stars, and modern  wide-field surveys [e.g., Sloan Digital Sky Survey (SDSS; \citealt{York00}), Pan-STARRS1 \citep{Kaiser10}, the Dark Energy Survey (DES; \citealt{Flaugher12}),  LSST] almost exclusively use broadband filters in order to facilitate extragalactic science goals.  

Recently, significant progress was made regarding the photometric estimation of \logg\ for stars with $4500$ K $<$ \teff\ $< 6750$ K, following the recognition that stellar brightness variations on timescales of several hours arise from granulation, which, in turn, correlates with surface gravity \citep{Bastien13}. This method, which produces a scatter of 0.06--0.10 dex in \logg\footnote{Throughout this paper we report \logg\ measurements in the cgs system. Thus, all references herein to \logg\ should be interpreted as $\log [g / ({\rm cm \; s}^{-2})]$, which we give in units of dex.} \citep{Bastien13}, requires high-quality ($\sim$0.01 millimag precision), high-cadence (every 30 min), monitoring from space-based telescopes, such as the \textit{Kepler} satellite \citep{Borucki10}. Wide-field, ground-based surveys will never achieve this precision, however, while high-cadence, high-quality \textit{Kepler}-like light curves are only available for a few hundred thousand stars that are restricted to a small number of specific sight lines. Given the proliferation of wide-field ($\ga{20}{,}000 \deg^2$), broadband, photometric surveys, our understanding of stellar evolution and the formation of the Milky Way would greatly benefit from the reliable determination of temperature, surface gravity, and metallicity for the hundreds of millions of stars observed by these surveys. 

Here, we present a new machine-learning framework and model capable of inferring the fundamental stellar atmospheric parameters, \teff, \logg, and \feh, from de-reddened photometric colors and time-domain observations alone. We train our algorithms using targets from our large ($\sim$9,000 sources), systematic spectroscopic survey of variable stars. For the sources in our survey we measure \teff, \logg, and \feh\ using well-established techniques that have been adapted from the SDSS survey. Our final models enable the precise measurement of these parameters for variable stars without (expensive) spectroscopic measurements.

\section{Systematic Spectroscopic Survey of Variable Sources}\label{survey_section}

To facilitate the construction of a large training set, we conducted a systematic, spectroscopic  survey of variable sources in \stripe, an equatorial, $\sim$315 deg$^2$ field that was repeatedly imaged  by the SDSS. Observations were conducted with Hectospec \citep{Fabricant05}, a multi-object spectrograph on the 6.5-m Multi-Mirror Telescope (MMT). 

\subsection{\stripe}

The SDSS repeatedly scanned an equatorial region of the southern Galactic cap, known as \stripe\ ($20^{\rm h}00^{\rm m} < \alpha_{\rm J2000.0} < 04^{\rm h}08^{\rm m}$, $-01^{\circ}16^{\rm m} < \delta_{\rm J2000.0} < 01^{\circ}16^{\rm m}$) during the first $\sim$9 yr of the survey. These repeated scans, conducted in each of the $ugriz$ filters, cover a wide range of galactic latitudes ($-15^{\circ} < b < -64^{\circ}$), and have enabled numerous time-domain studies. The $\sim$decade long, multi-color observations of \stripe, provide a superb testing ground for the eventual observations from LSST.

\subsection{Survey Design}

As a byproduct of a search for standard stars in \stripe, researchers at the University of Washington constructed a publicly available\footnote{Summary statistics and light curves of the variable candidates can be found on the UWVSC website: \url{http://www.astro.washington.edu/users/ivezic/sdss/catalogs/S82variables.html}.} variable source catalog (UWVSC; \citealt{Ivezic07,sesar07}). The UWVSC contains 67,507 unresolved, variable candidates with $g \le$ 20.5 mag, at least 10 observations in both the $g$ and $r$ bands, and a light curve with a root-mean-scatter (rms) $>$ 0.05 mag and $\chi^2$ per degree of freedom $>$ 3 in \textit{both} the $g$ and $r$ bands.
 
We adopt the UWVSC as the basis for our spectroscopic survey of variability for several reasons: (i) it is one of the largest existing catalogs of variable stars, (ii) it is the closest publicly-available analog to the data set that will ultimately be delivered by the LSST, and (iii) low galactic latitudes are included (see below).

\subsubsection{Target Selection}

Maximizing the efficiency of Hectospec requires a large density of targets ($\sim300 /\deg^2$). As a result, we elected to focus our survey on the lowest galactic latitudes in \stripe, the $\sim$25 deg$^2$ region with $300^\circ \le \alpha_{\rm J2000.0} \le 310^\circ$, which corresponds to galactic latitudes $-14.7^\circ \ge b \ge -24.6^\circ$. Hectospec fibers are positioned in a radial configuration extending inwards from the outer edge of the FOV. While this configuration allows the robotic fiber positioners to rapidly reconfigure the observational setup, a disadvantage of this design is that the radial fiber configuration results in a geometry where targets can conflict with one another requiring two fibers to cross, which is not possible. Thus, while 300 fibers are available for observations, in practice, typically only $\sim$175--200 science targets can be observed in a single pointing.

Given the dual goals of the survey to obtain a minimally-biased spectroscopic view of variability and to improve photometric classifiers of variable stars we employ additional selection criteria beyond the spatial location of the variable sources. We require all targets to have a mean \textit{observed} $r \leq 19$ mag, which should result in a SNR $\geq$ 10 for 10 min exposures. Of the 26,419 UWVSC sources within the spatial bounds of our survey, 14,994
have $r \leq 19$ mag. Our ability to characterize the variability of a given source improves as the number of observations increases, and so we also require all potential targets to have $\geq$ 24 observations in each of the $g$, $r$, and $i$ bands. This further culls the final target list to 9635 unique sources.

\subsubsection{Target Prioritization}

For each field to be observed, the Hectospec targeting software assigns fibers based on the user-supplied relative priority of each individual target. This scheme, which assigns fibers to as many sources in the highest priority category as possible before assigning as many fibers in the second highest category and so on, allows us to ensure that the brightest and best observed sources are the most likely to be observed. In the end, we adopted 11 levels of prioritization, which are summarized in Table~\ref{tbl-hecto_targs}. Generally speaking, we assigned higher rank (priority 1 corresponds to the highest rank) to brighter sources with more observations. 

We elevated the priority of targets of interest, sources that stand out regardless of their brightness or the total number of times they were observed. Three different categories were identified as high-priority: (i) high-amplitude sources, which we define as those having a median of absolute deviation (MAD), a robust measure of the scatter about the median, in the $r$ band $>$ 0.15 mag, (ii) sources that are likely periodic,\footnote{Periodicity was analyzed using a generalized Lomb-Scargle periodogram \citep{Lomb76,Scargle82,Zechmeister09} to analyze each source (see \citealt{richards11} for more details on our Lomb-Scargle periodogram implementation). Sources with a Lomb-Scargle power spectral density $>$ 16.5 were selected as likely periodic variables.} and (iii) sources with light curves that are consistent with the variability signature of quasars.\footnote{We identify sources with $\chi^2_{\rm QSO}/\nu > 5$ as sources similar to quasars (see \citealt{butler11} for a definition of $\chi^2_{\rm QSO}/\nu$).} This later group is of interest because quasars are difficult to find at low Galactic latitudes (e.g., \citealt{butler11}), yet they serve as ideal probes of the interstellar medium. Finally, we note that the 3  spectroscopically-confirmed quasars in the UWVSC that match our targeting criterion were excluded from the target list.

\begin{figure}
\centerline{\includegraphics[width=3.4in]{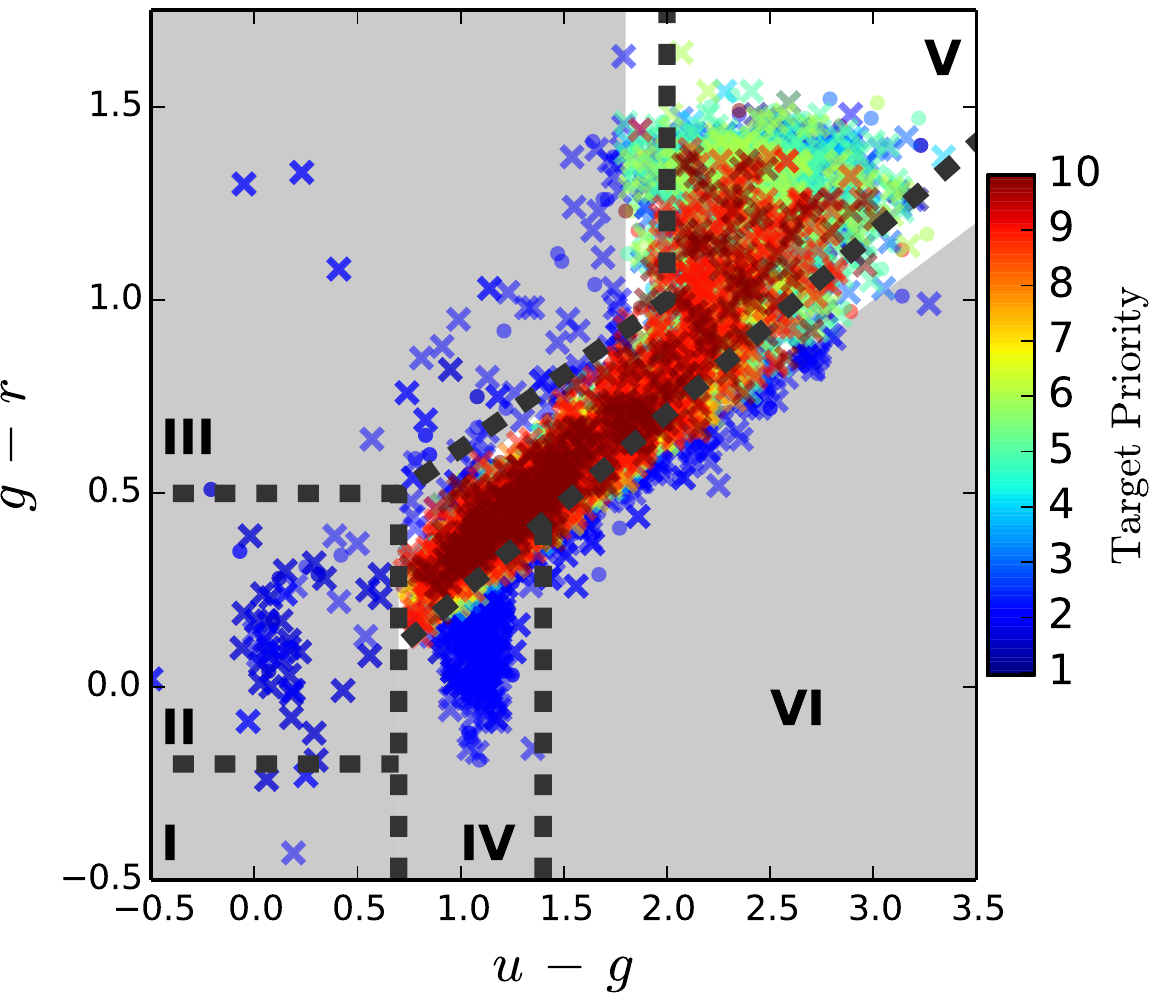}}
\caption[Color-color diagrams showing Hectospec targets]{$u \, - \, g$, $g \, - \, r$ color-color  diagram showing the $\sim$10k  potential Hectospec targets. Crosses show targets that were actually observed by Hectospec, while  the circles show targets that were not observed. The grey shaded area shows the region in color-color  space considered  outside the main stellar locus for targeting purposes. All sources in this region were assigned priority 1 or 2. The distribution of targets reasonably reflects the entire UWVSC sample (see \citealt{sesar07}), with the exception of low-redshift quasars, which are less prevalent due to the magnitude cuts and low Galactic latitude of the Hectospec fields. The six regions identified in \cite{sesar07} are outlined by the dashed lines. For each region, the variability is believed to be dominated by the following: \textbf{I} -- white dwarfs, \textbf{II}~--~low-redshift ($z \leq 2.5$) quasars, \textbf{III} -- white dwarf-M star binary stars,  \textbf{IV} -- RR Lyrae stars, \textbf{V} -- normal stars, \textbf{VI} -- high-redshift  ($z \geq 3$) quasars. Most of the high-amplitude variables are located off the main stellar locus.}
\label{fig:targ_CC}
\end{figure}

Each target was assigned a relative priority based on its brightness and total number of observations. Targets with priority 2.5 (see below) were repeated in the target list. The detailed criteria for the priorities we assigned to each target are as follows:
\begin{itemize}
    \setlength{\itemsep}{1pt}
    \setlength{\parskip}{0pt}
    \setlength{\parsep}{0pt}
    \item Priority 1 targets are those determined to be quasar-like, likely periodic variables, or  bright sources with either mean \textit{observed} $u$, $g$, or $r \le 15$ mag, or $i \le 14.8$ mag,  or $z < 14.7$ mag.\footnote{Magnitude cuts are brighter in the redder bands to prevent an unbalanced  selection of very-red objects, which are difficult to classify with low-resolution spectra (e.g., \citealt{lee08}). For all priorities below the same offsets apply, such that $ugriz \le m$ mag means  that the observed mean magnitude in $u$, $g$, or $r \le m$ mag, or $i \le m-0.2$ mag, or  $z \le m-0.3$ mag. A source only needs to be brighter than these limits in a single band for inclusion  in a given priority level.}   
    \item Priority 2 targets are those outside the main stellar locus in color-color space (see  Fig.~\ref{fig:targ_CC}) or those with $ugriz \le 16$ mag.
    \item Priority 2.5 targets are either extremely bright, $ugriz \le 14.6$ mag, or quasar-like, or  likely periodic variables, or high-amplitude variables. Note that inclusion as a Priority 2.5 target  is the only way a source could be added to the target catalog more than once.
    \item Priority 3 targets have been observed $\ge$30 times in the $r$ band and have $ugriz \le 17$ mag.
    \item Priority 4 targets have $ugriz \le 17$ mag.
    \item Priority 5 targets have been observed $\ge$30 times in the $r$ band and have $ugriz \le 18$ mag.
    \item Priority 6 targets have $ugriz \le 18$ mag.
    \item Priority 7 targets have been observed $\ge$30 times in the $r$ band and have $r \le 18.5$ mag.
    \item Priority 8 targets have $r \le 18.5$ mag.
    \item Priority 9 targets have been observed $\ge$30 times in the $r$ band and have $r \le 19$ mag.
    \item Priority 10 targets have $r \le 19$ mag.
\end{itemize}
While priority 2.5 sources were potentially observed twice, in practice this rarely happened since the observations were not complete for sources with priority 1. Furthermore, the repeated observations of a few targets provides a check of the systematic differences of observations made through different fibers. In total there were 10,129 total potential targets selected for Hectospec observations (see Table~\ref{tbl-hecto_targs}). Fig.~\ref{fig:targ_CC} shows the distribution of potential targets in a $u \, - \, g$, $g \, - \, r$ color-color ($ugr$ CC) diagram. The distribution of targets reasonably reflects that of \stripe\ as a whole (compare with Fig.~4 in \citealt{sesar07}), with one exception: a paucity of sources in region II.  The magnitude cuts significantly reduce the number of candidate quasars, which are additionally more difficult to identify at low Galactic latitudes.

\begin{deluxetable*}{crllllrrcc}
\tabletypesize{\footnotesize}
\tablecolumns{10}
\tablewidth{0pt}
\setlength{\tabcolsep}{2pt}
\tablecaption{Summary of Hectospec Target Priorities\label{tbl-hecto_targs}}
\tablehead{\colhead{Priority} & \multicolumn{5}{c}{Selection Criteria} & 
    \colhead{Targeted} & \colhead{Observed} &  & \colhead{\%}}
\startdata
1 & &\phantom{$\cap$}$ \; \; ugriz\le15$ & $\cup$~~QSO $\cup$ P && & 423 & 315 & ~ & 74.5 \\
2 & ($ugriz>15$ & $\cap \; \; ugriz\le16$) & $\cup$~~~(\textcolor{red}{$\oslash$}$ \ast $)  && & 1189 & 844 &~& 71.0 \\
2.5 & & \phantom{$\cap$}$ \; \; ugriz\le14.6$ & $\cup$~~QSO $\cup$ P & $\cup$ ${\rm MAD}_r>0.15$ & & 497 & 190 & ~ & 38.2 \\
3 & $ugriz>16$ & $\cap \; \; ugriz\le17$ & $\cap$~~~~~$\ast$ & $\cap \;\; N_r \ge 30$ & & 1092 & 750 & ~ & 68.7 \\
4 & $ugriz>16$ & $\cap \; \; ugriz\le17$ & $\cap$~~~~~$\ast$ & $\cap \;\; 24 \le N_r < 30$ & & 448 & 244 & ~ & 54.5 \\
5 & $ugriz>17$ & $\cap \; \; ugriz\le18$ & $\cap$~~~~~$\ast$ &  $\cap \;\; N_r \ge 30$ & & 1930 & 1222 & ~ & 63.3 \\
6 & $ugriz>17$ & $\cap \; \; ugriz\le18$ & $\cap$~~~~~$\ast$ & $\cap \;\; 24 \le N_r < 30$ & & 758 & 452 & ~ & 59.6 \\
7 & $ugriz>18$ & $\cap \; \; r\le18.5$ & $\cap$~~~~~$\ast$ &  $\cap \;\; N_r \ge 30$ & & 749 & 423 & ~ & 56.5 \\
8 & $ugriz>18$ & $\cap \; \; r\le18.5$ & $\cap$~~~~~$\ast$ & $\cap \;\; 24 \le N_r < 30$ & & 342 & 208 & ~ & 60.8 \\
9 & $ugiz>18$ & $\cap \; \; 18.5<r\le19$ & $\cap$~~~~~$\ast$ & $\cap \;\; N_r \ge 30$ & & 1872 & 911 & ~ & 48.7 \\
10 & $ugiz>18$ & $\cap \; \; 18.5<r\le19$ & $\cap$~~~~~$\ast$ & $\cap \;\; 24 \le N_r < 30$ & & 829 & 355 & ~ & 42.8 \\
\hline
Total & &&&& & 10129 & 5914 & ~ & 58.4 \\
\enddata
\vspace{-0.3cm}
\tablecomments{All targets are required to have 300$^\circ \le \alpha_{\rm J2000.0} \le 310^\circ$, 
         a mean {\it observed} $r$ band magnitude $\le$ 19 mag, and at least 24 observations in each of the 
         $g$, $r$, and $i$ bands. The selection criteria symbols mean the following: {\bf QSO} -- light 
         curve is consistent with being a quasar following the method of \cite{butler11}; {\bf P} -- light 
         curve shows strong periodicity; $\mathbf{{\bf MAD}_r}$${>}$\textbf{0.15} -- the 
         median absolute deviation 
         in the $r$ band is greater than 0.15 mag; $\mathbf{N_r {\ge} 30}$ -- there are 30 or more 
         observations in the $r$ band; ${\ast}$ -- the {\it de-reddenned} colors are consistent 
         with the stellar locus, this is roughly equivalent to region V in the $u-g$, $g-r$ color-color 
         diagram as defined in \cite{sesar07}, the precise boundaries are shown in Figure~\ref{fig:targ_CC}; 
         \textcolor{red}{${\oslash}$}~${\ast}$ -- colors are 
         outside the main stellar locus; \textit{\textbf{ugriz}} ${<}$ \textit{\textbf{m}} -- the 
         {\it observed} mean mag of the source is 
         brighter than $m$ mag in the $u$ or $g$ or $r$ bands, or brighter than $m-0.2$ in the $i$ band, 
         or brighter than $m-0.3$ mag in the $z$ band; 
         \textit{\textbf{ugriz}} ${>}$ \textit{\textbf{m}} -- same as the previous designation 
         except the sources are fainter than $m$. Lastly, note that the targets with priority 2.5 are, by
         definition, repeated elsewhere in the target list. This was done to provide a test of any 
         systematic issues associated with the reduction pipeline (see text).}
\end{deluxetable*}

Limited telescope access and fiber conflicts (see above) prevented the acquisition of spectra for each of the 10,129 targets. After optimizing fiber configurations over the 40 observation fields we used to cover the survey area, we could, at most, obtain 7038 total spectra; our actual yield was 5914 (see Table~\ref{tbl-hecto_targs}).

\subsection{Hectospec Observations}\label{sec:observations}

Hectospec is a 300 fiber multi-object spectrograph with a circular, 1$^\circ$ diameter field of view (FOV). Spectra were obtained using the 270 groove mm$^{-1}$ grating, which provides a dispersion of 1.2~\AA\ pixel$^{-1}$ and a resolution of $\sim$6.2~\AA\ full-width half-max, suitable for measuring basic stellar properties.

Hectospec is operated in service mode, and observations for our program were carried out during June and July 2011. Each field was to be observed with 2$\times$300 s exposures, which would allow for the rejection of cosmic rays and provide a SNR $\ge$ 10 even for the faintest targets. At the discretion of the service observers, the exposure times for a few fields were extended during non-photometric conditions (Table~\ref{tbl-hecto_obs}). In total, 33 of our 40 planned fields were observed, which yielded 5914 spectra of 5825 unique sources.

\begin{deluxetable*}{lrrrcccc}
\tabletypesize{\footnotesize}
\tablecolumns{8}
\tablewidth{0pt}
\tablecaption{Summary of Hectospec Observations\label{tbl-hecto_obs}}
\tablehead{\colhead{UT Date} & \colhead{Field} & \colhead{RA} & \colhead{Dec} & 
    \colhead{Exp.\ Time\tablenotemark{a}}  & \colhead{Airmass} & \colhead{Seeing} & 
    \colhead{$N_{\rm spec}$} \\
\colhead{} & \colhead{} & \colhead{(J2000.0)} & \colhead{(J2000.0)} & 
    \colhead{(s)} & \colhead{} & \colhead{($''$)} & \colhead{}}
\startdata
2011-06-08.441 & 3 & 20:16:38.47 & $-$00:50:03.32 & 600.0 & 1.185 & 0.65 & 198 \\
2011-06-08.456 & 5 & 20:06:50.02 & $+$00:46:59.29 & 600.0 & 1.175 & 0.65 & 201 \\
2011-06-09.391 & 17 & 20:28:29.76 & $+$00:49:04.43 & 600.0 & 1.243 & 0.86 & 190 \\
2011-06-09.376 & 8 & 20:13:19.60 & $+$00:47:10.78 & 600.0 & 1.256 & 0.86 & 194 \\
2011-06-10.328 & 23 & 20:23:02.66 & $-$00:46:10.10 & 600.0 & 1.566 & 0.91 & 186 \\
2011-06-10.343 & 24 & 20:19:19.46 & $-$00:46:29.35 & 600.0 & 1.438 & 0.91 & 196 \\
2011-06-10.358 & 25 & 20:18:42.43 & $+$00:46:53.25 & 600.0 & 1.328 & 0.91 & 210 \\
2011-06-10.373 & 26 & 20:14:04.20 & $-$00:47:16.61 & 600.0 & 1.281 & 0.91 & 205 \\
2011-06-10.389 & 27 & 20:16:57.74 & $-$00:02:43.31 & 600.0 & 1.232 & 0.91 & 184 \\
2011-06-10.410 & 29 & 20:11:25.30 & $-$00:50:06.78 & 600.0 & 1.199 & 0.77 & 181 \\
2011-06-10.424 & 30 & 20:13:00.47 & $-$00:05:46.31 & 600.0 & 1.178 & 0.77 & 197 \\
2011-06-10.439 & 31 & 20:11:11.16 & $+$00:44:17.84 & 600.0 & 1.167 & 0.77 & 198 \\
2011-06-10.454 & 32 & 20:09:20.93 & $+$00:00:31.04 & 600.0 & 1.186 & 0.77 & 195 \\
2011-06-10.468 & 33 & 20:08:17.09 & $-$00:44:33.51 & 600.0 & 1.218 & 0.77 & 196 \\
2011-06-11.347 & 1 & 20:20:47.66 & $+$00:09:52.50 & 600.0 & 1.387 & 1.11 & 159 \\
2011-06-11.318 & 34 & 20:08:45.75 & $+$00:46:24.12 & 600.0 & 1.517 & 1.11 & 203 \\
2011-06-11.333 & 38 & 20:05:09.17 & $-$00:43:20.15 & 600.0 & 1.425 & 1.11 & 197 \\
2011-06-12.354 & 10 & 20:35:48.40 & $-$00:15:04.47 & 600.0 & 1.403 & 1.01 & 147 \\
2011-06-12.368 & 12 & 20:38:19.51 & $-$00:47:04.24 & 600.0 & 1.349 & 1.01 & 142 \\
2011-06-12.382 & 13 & 20:30:01.50 & $-$00:42:20.61 & 600.0 & 1.270 & 1.01 & 156 \\
2011-06-12.396 & 14 & 20:30:59.07 & $+$00:45:14.06 & 600.0 & 1.214 & 1.01 & 180 \\
2011-06-12.411 & 15 & 20:04:11.70 & $+$00:16:40.64 & 600.0 & 1.175 & 1.01 & 178 \\
2011-06-12.427 & 16 & 20:34:01.51 & $+$00:40:20.90 & 900.0 & 1.172 & 1.01 & 169 \\
2011-06-12.447 & 18 & 20:28:37.06 & $-$00:00:15.05 & 900.0 & 1.176 & 1.01 & 164 \\
2011-06-12.276 & 2 & 20:15:40.00 & $+$00:32:00.15 & 600.0 & 2.075 & 1.01 & 166 \\
2011-06-12.293 & 4 & 20:07:29.05 & $-$00:27:51.98 & 900.0 & 1.758 & 1.01 & 177 \\
2011-06-12.310 & 6 & 20:21:19.85 & $-$00:18:10.56 & 600.0 & 1.671 & 1.01 & 167 \\
2011-06-12.324 & 7 & 20:32:28.04 & $-$00:14:29.42 & 600.0 & 1.602 & 1.01 & 143 \\
2011-06-12.339 & 9 & 20:37:43.70 & $+$00:41:47.41 & 600.0 & 1.490 & 1.01 & 150 \\
2011-06-14.290 & 19 & 20:26:28.18 & $-$00:42:38.26 & 600.0 & 1.908 & 1.01 & 170 \\
2011-06-14.305 & 20 & 20:24:48.79 & $+$00:03:39.71 & 600.0 & 1.679 & 1.01 & 166 \\
2011-06-14.275 & 37 & 20:04:25.10 & $+$00:49:48.15 & 600.0 & 1.854 & 1.01 & 179 \\
2011-07-08.427 & 21 & 20:25:25.79 & $+$00:48:02.16 & 840.0 & 1.252 & 0.89 & 170 \\
\enddata
\tablecomments{All observations were obtained with the 270 groves mm$^{-1}$ grating, 
    which provides spectroscopic coverage from $\sim$3700--9200 \AA.}
\tablenotetext{a}{The program called for 2$\times$300 s exposures for each field, 
    however, the service observers elected to increase the exposure times or take 3 exposures on nights 
    with partial cloud cover.}
\end{deluxetable*}

\subsection{Data Reduction}

The Hectospec observations were reduced using standard procedures: bias subtraction, flat fielding, bad-pixel masking, and cosmic-ray rejection were performed by the \texttt{specroad}\footnote{\texttt{specroad} was developed at the Harvard Center for Astrophysics, for more details see: \url{http://tdc-www.harvard.edu/instruments/hectospec/specroad.html}.} reduction pipeline.

In addition to standard processing, \texttt{specroad} performs a throughput correction for each fiber, followed by a correction to a known red-light leak in the detector (see \citealt{Fabricant05}) and a correction for absorption due to the atmospheric A and B bands. The final step of the pipeline estimates the redshift of each spectrum using the \texttt{IRAF}\footnote{\texttt{IRAF} is distributed by the National Optical Astronomy Observatory, which is operated by the Association of Universities for Research in Astronomy (AURA) under cooperative agreement with the National Science Foundation.} task \texttt{DOSKYXCSAO}, which cross-correlates the spectrum against several template spectra of stars, galaxies, and quasars.

All spectra were flux-calibrated using a spectrum of the spectrophotometric standard star BD$+$28 4211, taken on UT 2011 06 12. Visual inspection of the spectra at this stage revealed that several sources had a sharp kink around $\sim$8100~\AA\ with continua rising at redder wavelengths. This effect is due to the red-light leak in the Hectospec detector. While \texttt{specroad} has a red-light leak correction method, many spectra remain affected by this known systematic issue following the pipeline reduction. For bright blue sources, Hectospec observations are also affected by second-order scattered light redward of $\sim$7000 \AA, though the magnitude of this effect is significantly less than that of the red-light leak. 

We developed a custom procedure to correct the continua of our targets redward of 7000 \AA. First, the spectral energy distribution (SED) of each source is determined from the median magnitude in each of the $ugriz$ filters, and, for sources detected by the Two Micron All Sky Survey (2MASS; \citealt{skrutskie-2mass}), the single-epoch 2MASS $JHK_s$ measurements. A spectral match for each source is then identified based on a fit of the SED to the spectra in the Pickles Stellar Library \citep{pickles98}. A low-order spline is fit to the Hectospec spectral continuum redward of 7000 \AA, and a multiplicative factor is determined to warp the spline fit to the same shape as the Pickles star continuum. We fit splines of order $k = 1-7$, and use the Bayes Information Criterion (BIC) to select the optimal model of the Hectospec spectral continuum. Once the correction factor is determined, each Hectospec spectrum is warped so that the continuum reward of 7000 \AA\ matches that of the most photometrically similar Pickles star. This procedure leaves considerable uncertainty regarding the true continuum redward of 7000~\AA, however, this should not greatly affect the final results of this study. The SEGUE Stellar Parameters Pipeline (SSPP; see below) relies exclusively on spectral features blueward of 8000~\AA\ to determine stellar parameters, so warping the continuum redward of 7000~\AA\ does not significantly alter the output from the SSPP. To confirm this was the case, we artificially decreased the SNR of the spectra redward of 7000~\AA\ using several different prescriptions, including setting the SNR$ = 1$, and found that the SSPP output was not significantly altered.

\section{SSPP Estimates of \teff, \logg, and \feh}\label{sec:SSPP}

For full details on the SSPP procedures see \citet{lee08, lee08a}. Here, we provide a brief overview of the SSPP methodology. The SSPP relies on external measurements of the radial velocity (RV) of a star in order to shift all spectra to a zero-velocity rest frame. It has been shown that input RVs incorrect by as much as 200 \kms\ do not significantly alter the output of the SSPP. Following the shift to a common rest frame, the SSPP then measures line indices for several prominent stellar absorption features (e.g., H$\alpha$, H$\beta$, {Ca} {II} H\&K, {Na} {I}, etc.). To measure these indices, continuum fits are made both globally, over the entire spectrum, and locally, from a line-free region blueward of the absorption feature to a line-free region redward of the absorption feature. A specific line index is then calculated for each continuum-fitting method by integrating the continuum-normalized flux over a pre-defined wavelength interval. The calculated line indices, along with continuum-normalized spectra covering different wavelength ranges and the SDSS photometric colors, are fed to multiple parameter estimation methods (e.g., neural networks, synthetic spectral matching, Ca II K line index technique, etc.), which each provide an estimate of \teff, \logg, and \feh. Each of the estimation methods is tuned to apply only to stars in a restricted range of $g-r$ colors and SNR, over which the method is shown to be reliable. The individual measurements of \teff, \logg, and \feh\ are robustly combined to provide the final adopted values, and corresponding uncertainties, of the stellar parameters. The number of methods employed to produce the final adopted parameters is also returned. For high signal-to-noise ratio (SNR) spectra with 4500 K $\le T_{\rm eff} \le$ 7500 K, the SSPP measures \teff, \logg, and \feh\ with typical uncertainties of 157 K, 0.29 dex, and 0.24 dex, respectively \citep{lee08}. While processing spectra, the SSPP flags stars for which it cannot provide reliable estimates of the stellar parameters, such as very hot stars, white dwarfs and M giants. 

The spectra we obtained with Hectospec provide a good match with those obtained by SDSS, making the SSPP an ideal tool for estimating stellar parameters. SDSS spectra cover a wavelength range from 3800--9200~\AA\ with a resolving power of $R \sim$1800, while Hectospec covers 3700--9100~\AA\ with a resolving power of $R \sim$1000. We use a slightly adapted version of the SSPP, which accounts for the lower resolution of Hectospec as compared to SDSS, to estimate \teff, \logg, and \feh\ for each of the sources observed during our survey. The results from the SSPP are summarized in Table~\ref{tbl-sspp_hecto}.

We supplement our 5914 Hectospec spectra with 3121 additional SDSS spectra of stellar UWVSC sources. Each of these 9035 spectra were processed with the SSPP, and estimates of \teff, \logg, and \feh\ were obtained for 5994 of those sources. The remaining sources had some peculiarity (most often low SNR or \teff$ < 4000$ K) such that the SSPP could not provide estimates of the fundamental atmospheric properties.

\begin{deluxetable*}{lccrrcrrcrrc}
\tabletypesize{\small}
\tablecolumns{12}
\tablewidth{0pt}
\setlength{\tabcolsep}{2pt}
\tablecaption{SSPP Stellar Parameters for Hectospec Targets\label{tbl-sspp_hecto}}
\tablehead{\colhead{UW ID\tablenotemark{a}} & \colhead{P\tablenotemark{b}} & \colhead{flags\tablenotemark{c}} &
    \colhead{$T_{\rm eff}$} & \colhead{$\sigma(T_{\rm eff})$} & \colhead{$N_{T_{\rm eff}}$\tablenotemark{d}}&
    \colhead{$\log \; g$} & \colhead{$\sigma(\log \; g)$} & \colhead{$N_{\log \; g}$\tablenotemark{d}}&
    \colhead{$[{\rm Fe/H}]$} & \colhead{$\sigma([{\rm Fe/H}])$} & \colhead{$N_{[{\rm Fe/H}]}$\tablenotemark{d}} \\
    \colhead{} & \colhead{} & \colhead{} & \colhead{(K)} & \colhead{(K)} & \colhead{} & \colhead{(dex)} & 
    \colhead{(dex)} & \colhead{} & \colhead{(dex)} & \colhead{(dex)} & \colhead{}}
\startdata
4172970 & 4 & nnnnn & 5499.7 & 94.7 & 7 & 3.94 & 0.18 & 6 & 0.01 & 0.12 & 5 \\
4583821 & 5 & Nnnnn & 4487.8 & 120.3 & 3 & 4.19 & 0.26 & 2 & -0.95 & 0.08 & 1 \\
4651452 & 2 & NnBnX & -9999.0 & -9999.0 & 0 & -9999.00 & -9999.00 & 0 & -9999.00 & -9999.00 & 0 \\
4777216 & 9 & nnnnX & -9999.0 & -9999.0 & 0 & -9999.00 & -9999.00 & 0 & -9999.00 & -9999.00 & 0 \\
5302673 & 9 & nnnnn & 5986.7 & 161.8 & 4 & 3.93 & 0.79 & 2 & -0.76 & 0.09 & 1 \\
\enddata
\vspace{-0.3cm}
\tablecomments{For sources where the SSPP was unable to measure \teff, \logg, or \feh, default values 
    of $-$9999 were adopted, as is shown in the second and third rows of this table. Only the first five sources are presented 
    here as an example of the form and content of the complete table. The full table, containing all 5914 sources observed 
    by Hectospec, is available online.}
\tablenotetext{a}{Source ID in the UWVSC.}
\tablenotetext{b}{Target Priority for Hectospec observations.}
\tablenotetext{c}{Analysis flags returned by the SSPP (see \citealt{lee08}).}
\tablenotetext{d}{Number of methods used by SSPP to derive final (adopted) parameters.}
\end{deluxetable*}

\section{Characterizing the Spectroscopic Sample of Variable Stars}

\begin{figure}[b]
    \centerline{\includegraphics[width=3.4in]{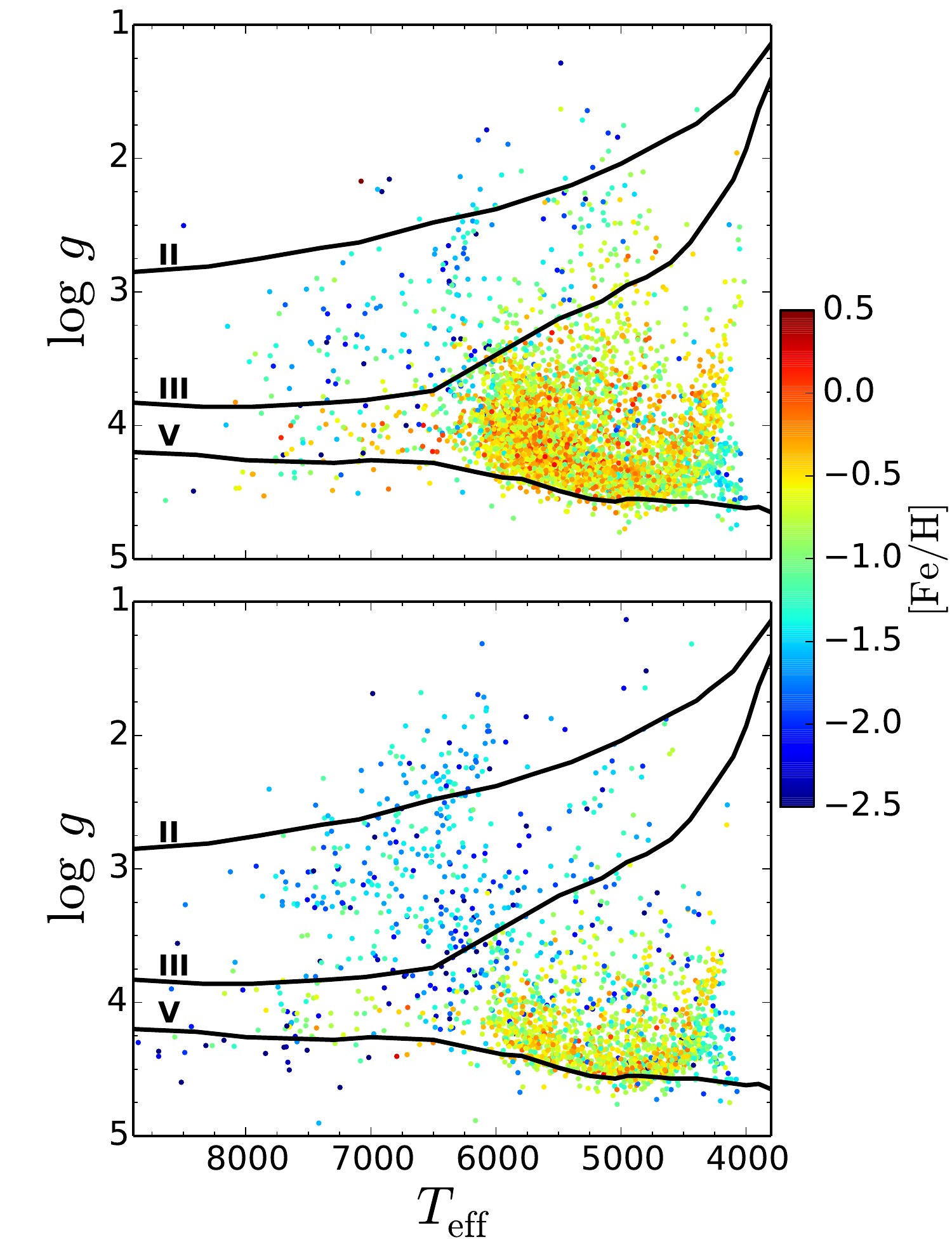}}
    \caption[\teff\ against \logg\ for stellar UWVSC sources]{\logg\ versus \teff\ for all UWVSC sources with atmospheric parameters measured by the SSPP. 
\textit{Top}: Sources with Hectospec spectra. \textit{Bottom}: Sources with SDSS spectra. Sources are color coded by their measured metallicity, as traced by \feh. For reference, the approximate locations of the dwarf, giant, and supergiant luminosity classes are shown via solid lines and marked V, III, and II, respectively. Note that the majority of variable stars in our sample are cool (\teff\ $\leq 6000$ K) with dwarf-like surface gravity (\logg\ $\geq 3.75$). The apparent trend of \feh\ against both \teff\ and \logg\ is a by-product of the magnitude-limited observations obtained by SDSS.}
    \label{fig:loggVteff}
\end{figure}

In Fig.~\ref{fig:loggVteff}, we show the distribution of \logg\ against \teff\ for all sources in our sample with SSPP estimates of these parameters. Also shown are the loci of dwarf, giant, and supergiant stars, denoted by their respective luminosity classes, V, III, and II. These locations, which are based on the tabulation of \cite{Straizys81}, are only approximate and serve as a rough guide for any individual source. From Fig.~\ref{fig:loggVteff} it is immediately clear that our sample of variable stars has few giants and supergiants. For a magnitude-limited survey it is not obvious that this should be the case, since giants are several mag brighter than dwarfs. It is also interesting that most sources have temperatures between $\sim$4000--6000 K. The pile up on the red end of this range is artificial and the result of the SSPP, which does not provide temperature estimates for stars cooler than 4000 K. From Fig.~\ref{fig:loggVteff} we conclude that the majority of variable stars are G and K dwarfs, with the caveat that a significant population of even cooler M type stars may constitute a significant fraction of the observed variables (see e.g., \citealt{Basri11}). 

The sources shown in Fig.~\ref{fig:loggVteff} are color coded via their SSPP measured metallicity. There is a general trend for warmer, low-gravity stars to have lower metallicity than the cooler, dwarf stars. We interpret this effect to be a result of the magnitude-limited observations of \stripe: low-metallicity dwarf stars in the halo are too faint to be detected, which biases the cooler stars to solar-like \feh. SDSS probes a larger volume for giant stars, and given the fixed area of \stripe, this means that halo giants will outnumber those found in the disk, biasing low-surface-gravity sources toward lower metallicities. 

We characterize the photometric variability of every UWVSC source in each of the $ugriz$ filters via the 66 light-curve features\footnote{In machine-learning parlance a ``feature'' is a real-numbered or categorical metric describing a source. The features listed in \citet{Richards12a} are based on either the time-series input or physical photometric colors of the source.} defined in \citet{Richards12a}. In total there are 334 features (66 for each of the $ugriz$ filters, as well as the 4 SDSS colors), that the machine learning model can use to map photometric variability to \teff, \logg, and \feh.

\begin{figure}
    \centerline{\includegraphics[width=3.5in]{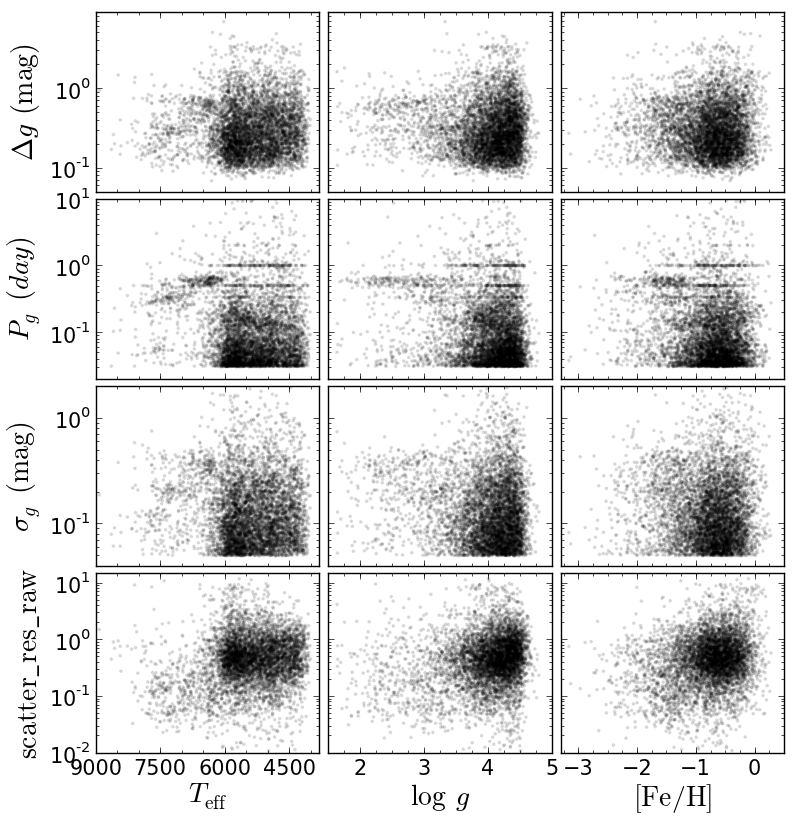}}
    \caption[Light-curve features against atmospheric parameters (1)]{The most  important light-curve features for automated variable-star classification are plotted against \teff, \logg, and \feh, as determined by the SSPP. In each case there is a single dominant clump, which consists primarily of main-sequence G and K dwarfs.}
\label{fig:feats1}
\end{figure}

\begin{figure}
    \centerline{\includegraphics[width=3.5in]{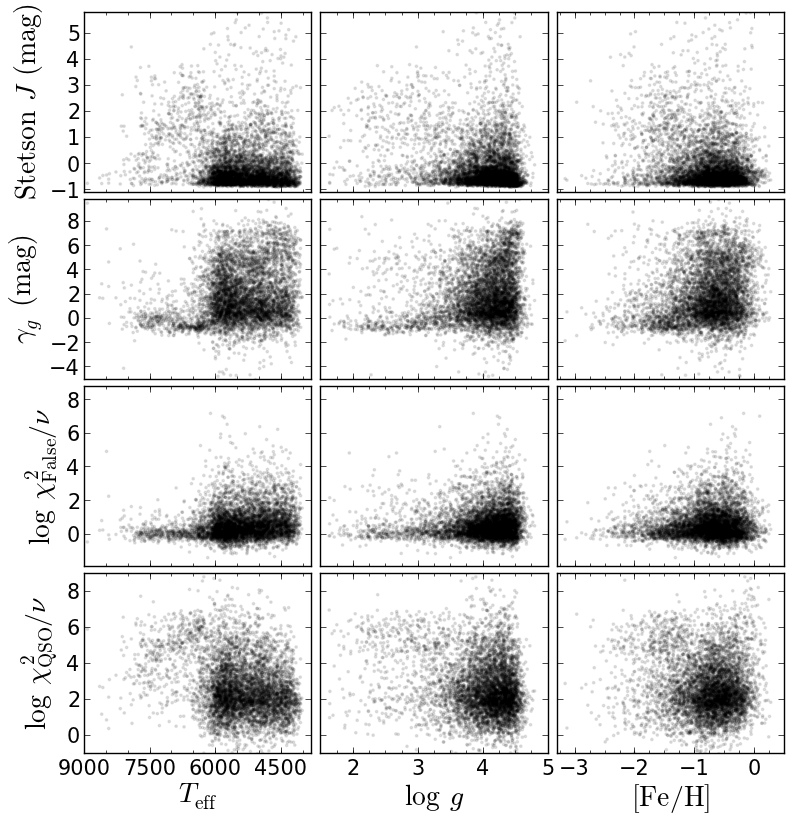}}
    \caption[Light-curve features against atmospheric parameters (2)]{Same as Fig.~\ref{fig:feats1}.}
    \label{fig:feats2}
\end{figure}

In Figures~\ref{fig:feats1}~and~\ref{fig:feats2}, we show the fundamental atmospheric parameters plotted against eight of the most important light-curve features for automated variable star classification, as determined in \cite{Richards12a}. The features, each measured from $g$ band observations, are: the amplitude of variations $\Delta g$, the best-fit period $P_g$, the standard deviation $\sigma_g$, the MAD of the Lomb-Scargle residuals divided by the MAD of the raw light curves,  \texttt{scatter\_res\_raw} (see \citealt{Dubath11}), the Stetson variability index $J$ (see \citealt{Stetson96}), the light-curve skewness $\gamma_g$, and the quasar and non-quasar variability metrics, $\chi^2_{\rm QSO}/\nu$ and $\chi^2_{\rm False}/\nu$, respectively (see \citealt{butler11}). If it is possible to estimate \teff, \logg, and \feh\ from light curves alone, then one would expect that at least some of the important light curve features correlate with the atmospheric parameters. From Figures~\ref{fig:feats1}~and~\ref{fig:feats2} it is clear that no obvious one-to-one correlations exist between the light-curve features and the stellar parameters. There are, however, some clear clumps that emerge from the data. For instance, RR Lyrae variables stand out as the warm (\teff~$\approx 7000$ K), low surface gravity (\logg~$\approx 2.5-3$), low metallicity (\feh~$\approx -2$) sources with $\gamma_g \approx -0.4$, $\chi^2_{\rm False}/\nu \approx 1$, and $P_g \approx 0.7$ d.\footnote{It has been shown that RR Lyrae stars cluster around these values of $\gamma_g$, $\chi^2_{\rm False}/\nu$, and $P_g$ \citep{sesar07,butler11}.} The existence of these clumps suggest that higher-dimensional models may be capable of parsing the multidimensional light curve feature space in order to predict fundamental stellar parameters.

\section{Construction of the Random Forest Regression Models}

\subsection{Pruning the Training Set}
In order to maximize the efficacy of our model, it is essential that the training set contain only sources with reliable estimates of \teff, \logg, and \feh. While the SSPP returns a set of internal flags designed to identify sources with unreliable parameter estimates, some sources flagged as ``nnnnn,'' meaning no flags were raised while processing, clearly have incorrect estimates. For instance, within our sample four ``nnnnn'' sources were identified as extremely metal poor (EMP) stars with \feh\ $< -3$ dex. Visual inspection of these sources shows that none of them are genuine EMP stars: two are white dwarfs, for which the SSPP is known not to perform well \citep{lee08}, one has an unusual continuum that is suggestive of an M star binary, and the fourth is a borderline candidate EMP star with low SNR ($=$ 10.9). It has been shown that some sources with SSPP estimates of \feh\ just below $-$3 dex are not in fact EMP stars \citep{Aoki13}. 

In order to further cull the training set to exclude sources with unreliable parameter estimates, we visually examined the SSPP diagnostic plots for each star in our spectroscopic sample. In addition to providing the SSPP estimates, these diagnostics display the observed spectrum plotted over a model spectrum of a star with the adopted SSPP parameters. These plots enable a quick visual analysis to determine the reliability of the estimated parameters. While visually inspecting each of the 9035 spectra, one of six visual inspection flags were assigned to each source: ``n'' -- spectrum appears normal, SSPP estimates are valid; ``X'' -- no SSPP parameters were estimated (typically due to SNR $<$ 10); ``Q'' -- spectrum shows broad emission lines consistent with a quasar; ``F'' -- parameters are estimated but the model spectrum clearly does not match the observed spectrum; ``M'' -- the star shows clear evidence for TiO and VO absorption, consistent with an M type star; and ``C'' -- the model and observed spectrum are well matched redward of $\sim$4500 \AA, but poorly matched in the spectral region around Ca II H\&K. Virtually all sources with visual examination flag ``C'' are late-type stars (\teff\ $< 5000$ K), with low SNR in the blue portion of the spectrum, which leads to the poor match between the observed and model spectra around Ca II H\&K. Based on the quality of the match between these spectra redward of 4500 \AA, we consider the estimates for these sources reliable. In the end, we adopt all sources with visual examination flags of either ``n'' or ``C,'' a total of 5881 sources, as the training set for our machine learning model.

\subsection{Random Forest Regression}

There are many machine-learning methods that can be used to perform supervised regression, including:  artificial neural networks, support vector machines, decision trees, and random forest, which have all been  successfully applied to large multidimensional datasets (see \citealt{Hastie09} for detailed examples of  the application of these tools). We employ the use of random forest regression~(RFR),  which is both fast and easy to interpret. Additionally, random forest has been  shown to be the optimal machine-learning method for a variety of astrophysical problems (e.g.,  \citealt{richards11, Dubath11, Brink13, Morgan12}). A detailed description of the random forest algorithm can be found in \citet{Breiman01}. Briefly, the random forest method aggregates the results from several decision  trees to provide a low-bias, low-variance estimate of the properties of interest. In particular, at each node of the tree the new splitting parameter can only be selected from a random subset of \texttt{mtry}  features in the entire feature set. For the case described here, after hundreds of trees have been constructed, each with different structure, the output from each of those trees is averaged to provide a  robust estimate of \teff, \logg, and \feh. Finally, we note that each parameter estimate comes from a separate model, one for each of \teff, \logg, and \feh. 

We adopt the root mean-square error (RMSE) as the figure of merit (FoM) for selecting the parameters of one model over another. When applied to the training set, the RMSE is defined as:  
    $$\textrm{RMSE} = \sqrt{\frac{1}{n}\sum_i (y_i - x_i)^2},$$ %
where $n$ is the total number of objects in the training set, $y_i$ is the predicted value of the property  of interest, and $x_i$ is the spectroscopic value of the property of interest. The splitting parameter of each non-terminal RFR node is optimized to minimize the RMSE, making this a natural choice for the FoM. The RMSE for the entire training set is measured using $k$-fold cross validation (CV). In $k$-fold CV, 1/$k$ of the training set is withheld during model construction, and the remaining $1 - 1/k$ fraction of the training set is used to predict the parameters of interest for the withheld data. This procedure is then repeated $k$ times, resulting in every training set source being withheld exactly once, so that predictions are made for each source in the training set enabling a measurement of the RMSE.

\subsection{Improving Stellar Parameter Estimates with Time-Domain Information}

We endeavor to improve stellar parameter estimates from photometric observations by supplementing SDSS photometric colors with time-domain information. While the acquisition of photometric light curves used to be very costly, similar to spectroscopy, recent advances in both the construction of large-format charge-coupled devices (CCDs) and in computational processing and data storage have enabled an unprecedented exploration of the time domain over wide fields. Currently, several surveys repeatedly image $>$10,000 deg$^2$ (e.g.,  \citealt{Pojmanski01,law09,Larson03,Keller07}), with several more (e.g., \citealt{Tonry11}) planned prior to LSST. 

We begin by examining the utility of supplementing photometric colors with light-curve features in order to estimate fundamental stellar parameters. Since many variable star classes correspond to specific locations within the Hertzsprung-Russell diagram (e.g., RR Lyrae stars, Cepheid stars, Mira variables; see \citealt{Eyer08}), it seems reasonable to expect that light-curve features, in addition to colors, could, at the very least, improve estimates of \logg. Furthermore, the observed correlation between metallicity and periodic-light-curve features for some variables (e.g., Cepheids, see \citealt{Klagyivik13}), suggests that time-domain observations can improve estimates of \feh\ for at least some stars. Note, however, we do not directly classify variables as part of the present framework. One potential complication for our approach is binarity: many of the variables in the UWVSC are actually two stars orbiting each other, which leads to a more complicated spectrum than those from a single star. In practice, however, the light from many of these systems will be dominated by a single star, in which case our method should remain valid. Alternatively, if the flux contribution is comparable from the two sources, then they have similar mass, and hence similar \logg, while the metallicity will be identical as the stars were formed from the same molecular cloud. Thus, we do not expect binaries to significantly alter the results from this study. 

As an initial test to determine whether light-curve features can be used to estimate \teff, \logg, and \feh\ for variable sources, we construct three RFR models for each of these three atmospheric parameters. The feature set for the first model contains only the median observed SDSS photometric colors ($u - g$, $g - r$, $r-i$, and $i - z$), which have been corrected for reddening according to the dust maps of \citep{Schlegel98}.\footnote{Below we will show that photometric colors are important for estimating \teff, \logg, and \feh. We use reddening corrections from SDSS, which work best at high galactic latitudes. The \citeauthor{Schlegel98} dust maps are known to be less accurate in the Galactic plane, and thus, the efficacy of our models has not been tested in this region.} The feature set for the second model contains only the 330 (66 features from each of the $ugriz$ bands) light-curve features described in the text, while the third model uses all 334 (both colors and light-curve) features. In some cases, correlated features can hurt the performance of random forest models (see e.g., \citealt{Dubath11}). In order to avoid correlations to the features measured for each of the five SDSS filters, our feature set includes the 66 features from \citep{Richards12a} measured on the $g$ band light curves, along with the difference between the values of these features in the remaining filters. For instance, rather than including the amplitude of variations in each filter $\Delta_{\rm filter}$, our feature set includes $\Delta_g$, $\Delta_u - \Delta_g$, $\Delta_g - \Delta_r$, $\Delta_r - \Delta_i$, and $\Delta_i - \Delta_z$.  

\begin{figure*} 
    \centerline{\includegraphics[width=6in]{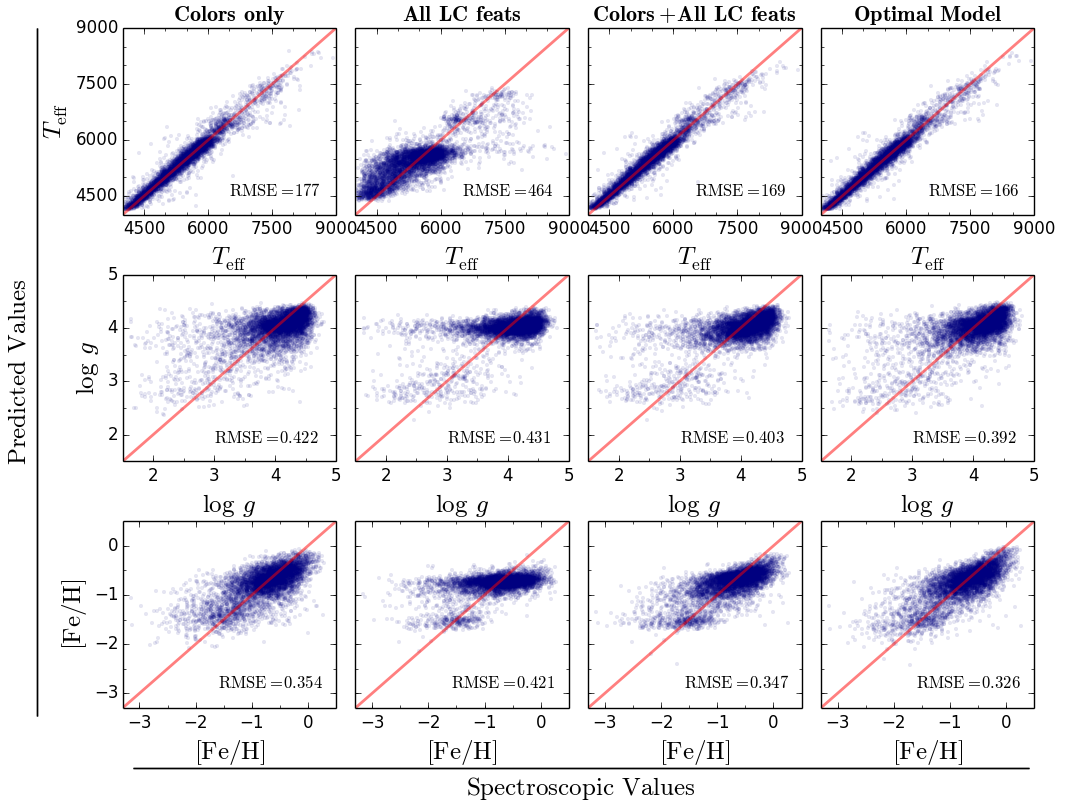}}   
    \caption[Random forest regression using photometric colors and light-curve features]{Random forest regression predictions vs.\ spectroscopic values measured by the SSPP. The top row shows the results for \teff, the middle row shows \logg, and the bottom row shows \feh. The columns show models constructed using different feature sets: the first column shows models fit to only SDSS colors, the second column shows models fit to the 330 light-curve features only, the third column shows models fit using both SDSS colors and light-curve features, and the fourth column shows the final optimized model. For comparison, the RMSE of \textit{the complete training set} for each model is shown. Note that the RMSE values quoted in Fig.~\ref{fig:final_model} refer to the ``pristine'' subsample of the training set (see \S\ref{sec:opt_CV_results}), which is why those values differ from the ones quoted here.} \label{fig:model_compare} 
\end{figure*}

The results of this initial test, as well as those for our final optimized model, are shown in Fig.~\ref{fig:model_compare}. From the second column of Fig.~\ref{fig:model_compare}, it is obvious that the 330 light-curve features alone are a poor predictor of \teff, \logg, or \feh. The first column shows colors may be used to estimate \teff\ with a scatter similar to that produced from SSPP measurements of actual spectra. Colors-only estimates of \logg\ and \feh\ are significantly worse than what can be gleaned from low-resolution spectra, however. The third column shows that using both colors and light-curve features provides modest gains relative to models trained using only color information. These models have not been optimized, however, and with 330 light-curve features and a median of only 33 observations per source it is likely that these models have been over-fit. As the fourth column shows, reducing the total number of features and optimizing the RFR parameters provides the best estimates for \teff, \logg, and \feh, as we discuss in more detail below.  

The importance of the many model construction decisions that a scientist must make can plainly be seen in Fig.~\ref{fig:model_compare}: when using light-curve features and excluding color information to predict \feh, RFR produces very biased results in which almost all sources are predicted to have \feh\ $\approx -0.8$ dex. Using this model, it would be virtually impossible to identify newly observed low-metallicity stars. Below we describe how we prune the feature sets used to predict \teff, \logg, and \feh, and how we adjust the RFR tuning parameters to construct the optimal machine-learning model.

\section{Optimizing the Models}

\subsection{Feature Selection}

For variable-star classification, each of our adopted 66 light-curve features provides useful information for discriminating between the variability classes \citep{Richards12a}. Nevertheless, the use of all 330 light-curve features adds noise to the model and reduces its overall accuracy. The inclusion of too many features can hinder the performance of the model, needlessly making it more complicated while increasing the likelihood of over-fitting the data.

Feature selection is a challenging problem: there is an exceedingly large number of combinations that include a subset of the 334 total features. We wish to determine which subset $s$ of the 334 features produces the best model to predict stellar parameters. Searching over all possible subsets is computationally intractable as it would require $\sum_{s = 2}^{334} (^n_s) \approx 10^{100}$ RFR models.\footnote{Random forest methods require at least two features to construct a meaningful decision tree.}

We simplify the model selection process using the well-known forward-feature selection approximation method \citep{Guyon03}. The forward-feature selection method begins with an empty feature set and iteratively adds features one at a time, selecting the feature that best improves the model, as judged by the FoM, at each step. Due to the randomness of RFR, this procedure is repeated five times, and the features with the highest median importance are selected as the final feature set for our model. 

Before proceeding with forward feature selection for each of our three models, we must identify one feature that is automatically included in the final feature set. One of the advantages of random forest, over other algorithms, is that during the model construction process the relative importance of each feature is naturally and automatically measured, since a subset of features are excluded as splitting parameters in each non-terminal node of the tree \citep{Breiman01}. Thus, we perform RFR using the full feature set, and adopt the most important feature determined by the random forest algorithm as the initial feature for forward selection. For each of the \teff, \logg, and \feh\ models the most important feature is $g - r$. From there, we aim to only add features that improve the FoM of the model. To do so, we calculate the cross-validated FoM of all possible models with a single feature added to $g - r$, and select the model with the smallest RMSE. We repeat this process until a model with 50 features has been constructed. Truncating the model at 50 features significantly reduces the computation time relative to forward selection over all 334 features. While the selection of 50 features is arbitrary, in practice, this choice extends the procedure well beyond the optimal number of features as measured by the FoM (see below), meaning that it has no effect on our final models. The forward selection procedure requires the creation of $\sum_{i = 1}^{50} 334 - i = 15,\!425$ RFR models, which, though large, is tractable, unlike a complete search over all possible model combinations.

\begin{figure}
    \centerline{\includegraphics[width=3.5in]{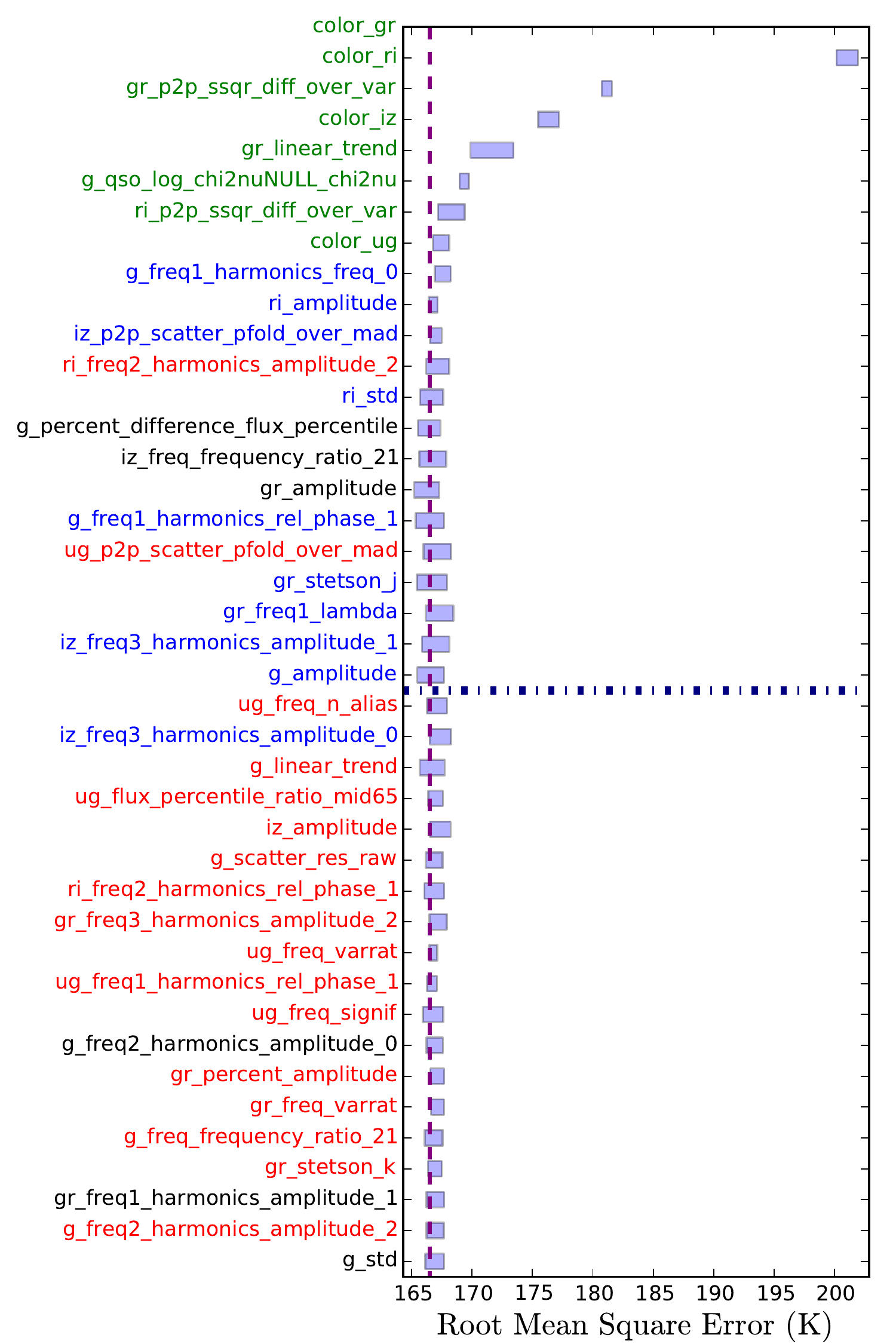}}
    \caption[Forward feature selection for \teff]{Results of the forward feature selection process for \teff. We begin by selecting $g - r$ (\texttt{color\_gr} in the figure) and iteratively add the feature that most improves the regression model as measured by the improvement in the FoM. Boxes show the cross-validated range of RMSE following the addition of the feature to the model. The vertical dashed line shows the smallest median RMSE. Features above the dash-dot line, which is defined by the first feature with larger median RMSE than the previous feature, are those that are selected for the optimal feature set. The procedure was rerun 5 times and the feature names are color coded according to the number of times they were selected in the optimal feature set: 0 (black), 1 (red), 2 (blue), 3+ (green). For brevity, only the first 40 selected 
features are shown.}
    \label{fig:teff_ff}
\end{figure}

\begin{figure}
    \centerline{\includegraphics[width=3.5in]{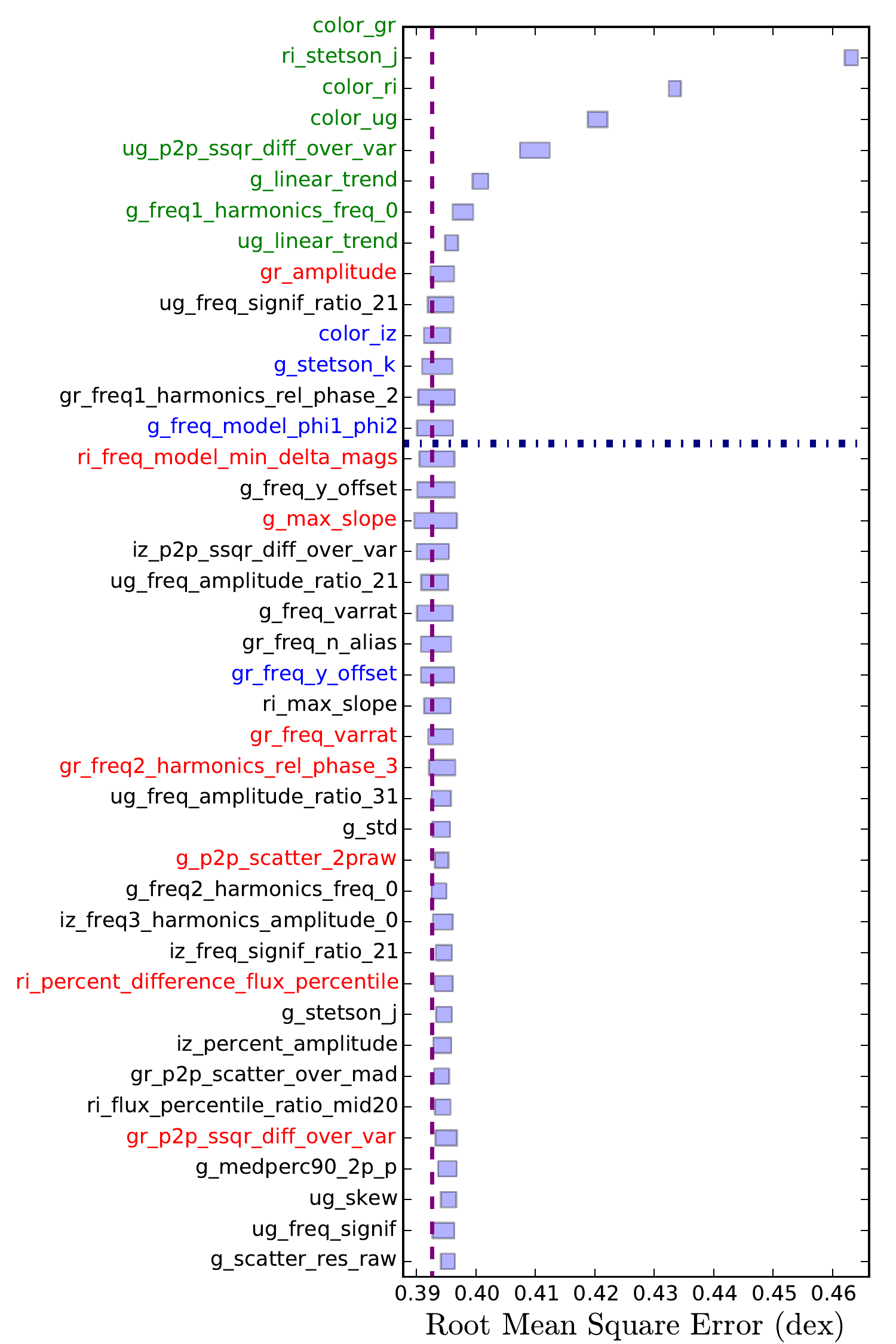}}
    \caption[Forward feature selection for \logg]{Results of the forward feature selection process for \logg. The explanation for this figure is the same as Fig.~\ref{fig:teff_ff}.}
    \label{fig:logg_ff}
\end{figure}

\begin{figure}
    \centerline{\includegraphics[width=3.5in]{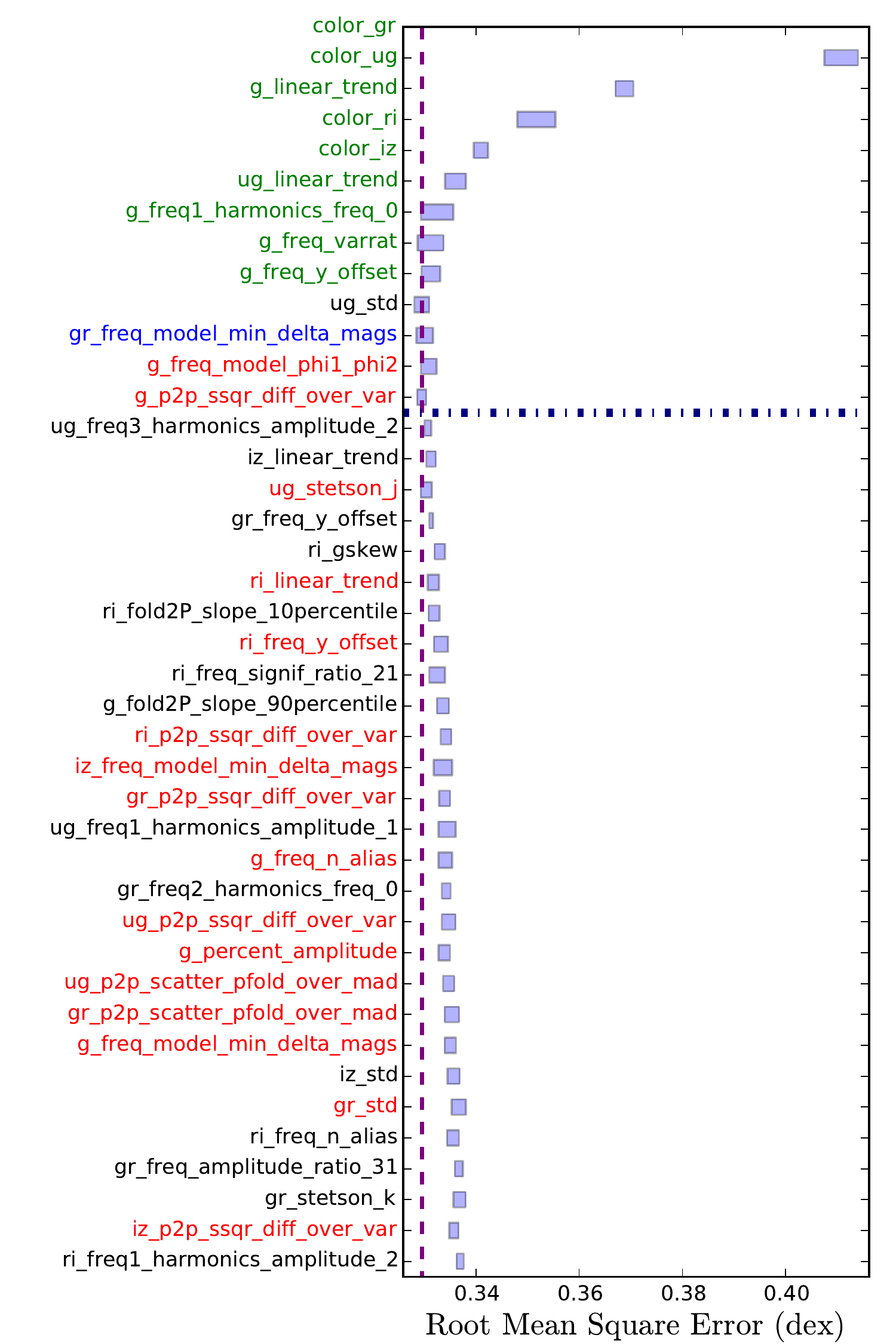}}
    \caption[Forward feature selection for \feh]{Results of the forward feature selection process for \feh. The explanation for this figure is the same as Fig.~\ref{fig:teff_ff}}
    \label{fig:feh_ff}
\end{figure}

Figures~\ref{fig:teff_ff},~\ref{fig:logg_ff},~and~\ref{fig:feh_ff} show the results from the forward feature selection process for \teff, \logg, and \feh, respectively. There is an improvement in the performance of the RFR model when the feature set is pruned. 

An important thing to note from Figures~\ref{fig:teff_ff}--\ref{fig:feh_ff} is that the feature selection method is robust, since the ordering of features does not change significantly from run to run. The final selected features are selected in most of the runs, while only a few features are selected just once or twice. The occasional inclusion of a low-ranking feature in an individual run is the result of correlations between the features.

\subsection{Tuning the Model}

Random-forest methods feature three important tuning parameters: (i) \texttt{ntree}, the total number of decision trees used to construct the forest, (ii) \texttt{mtry}, the number of features that are used as potential splitting criterion in each non-terminal node of the tree, and (iii) \texttt{nodesize}, the minimum number of training set objects, meaning further splitting is not allowed, in a tree's terminal nodes. To optimize the random forest tuning parameters, we perform a grid search over \texttt{ntree}, \texttt{mtry}, and \texttt{nodesize}.

For RFR the ``rule-of-thumb'' values are \texttt{mtry} = $\sqrt{n}$, where $n$ is the total number of features used in the model, and \texttt{nodesize} = 5. \texttt{ntree} is specified by the user, during forward feature selection we selected \texttt{ntree} = 100. Broadly speaking, adjusting the tuning parameters adjusts the smoothness of the model, as they affect the complexity of the random forest. Now that we have engineered an optimal feature set, we aim to determine which values of the training parameters will produce the optimal RFR models to predict \teff, \logg, and \feh.

\begin{figure*}
    \centerline{\includegraphics[width=4.5in]{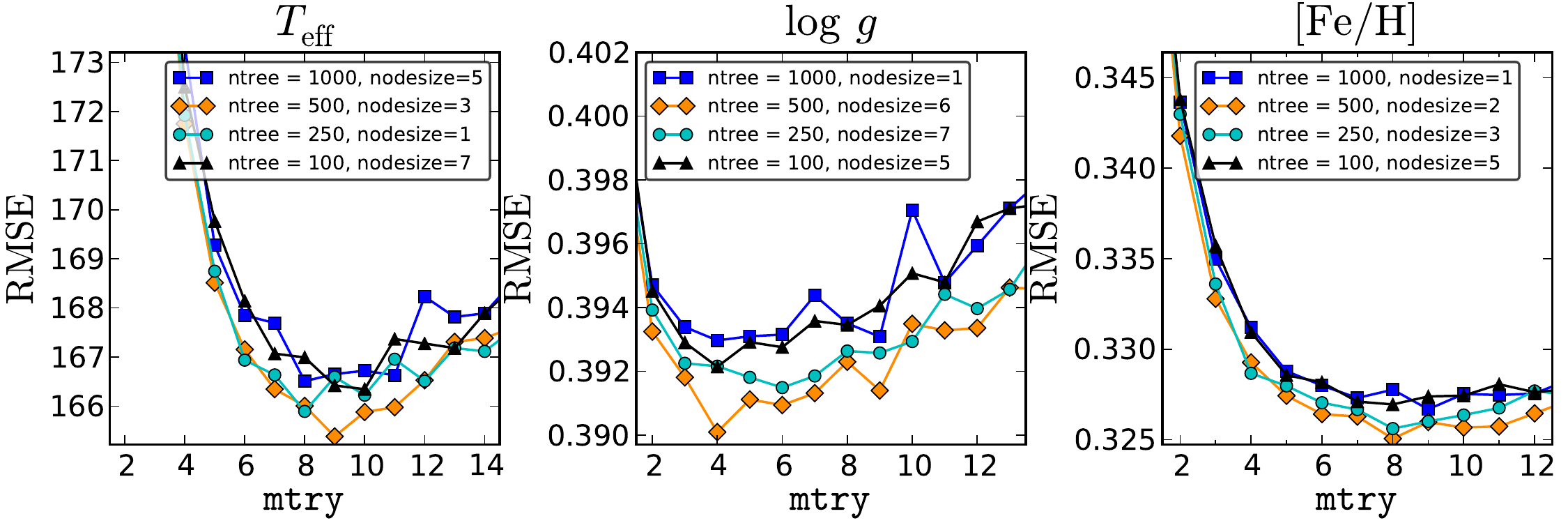}}
    \caption[Optimal random forest model selection]{Results from a grid search over the three random forest parameters: \texttt{ntree}, \texttt{mtry}, and \texttt{nodesize}. The models are run a total of three times, and the displayed results show the average RMSE. Note that the models are not strongly sensitive to the choice of tuning parameters. For \teff, the optimal model is: $\texttt{ntree} = 500$, $\texttt{mtry} = 9$, and $\texttt{nodesize} = 3$. For \logg, the optimal model is: $\texttt{ntree} = 500$, $\texttt{mtry} = 4$, and $\texttt{nodesize} = 6$. For \feh, the optimal model is: $\texttt{ntree} = 500$, $\texttt{mtry} = 8$, and $\texttt{nodesize} = 2$.}
    \label{fig:model_select}
\end{figure*}

To optimize the random forest tuning parameters, we perform a grid search over \texttt{ntree}, \texttt{mtry}, and \texttt{nodesize}. Selected results from this grid search are shown in Fig.~\ref{fig:model_select}. Generally, we find that the precise choice of tuning parameters does not significantly affect the output of the models, as measured by the FoM. The behavior of the models as \texttt{mtry} is adjusted is typical for non-parametric classifiers: small values of \texttt{mtry} are over-smoothed, high-bias and low-variance models, while large values of \texttt{mtry} lead to under-smoothed, low-bias and high-variance models. Following forward-feature selection and the tuning of the RFR model parameters, we have constructed an optimized model for the prediction of \teff, \logg, and \feh\ from photometric light curves. 

\section{Results}\label{sec:final_results}

\subsection{Optimized CV Results}\label{sec:opt_CV_results}

\begin{figure*}[t]
    \centerline{\includegraphics[width=6.2in]{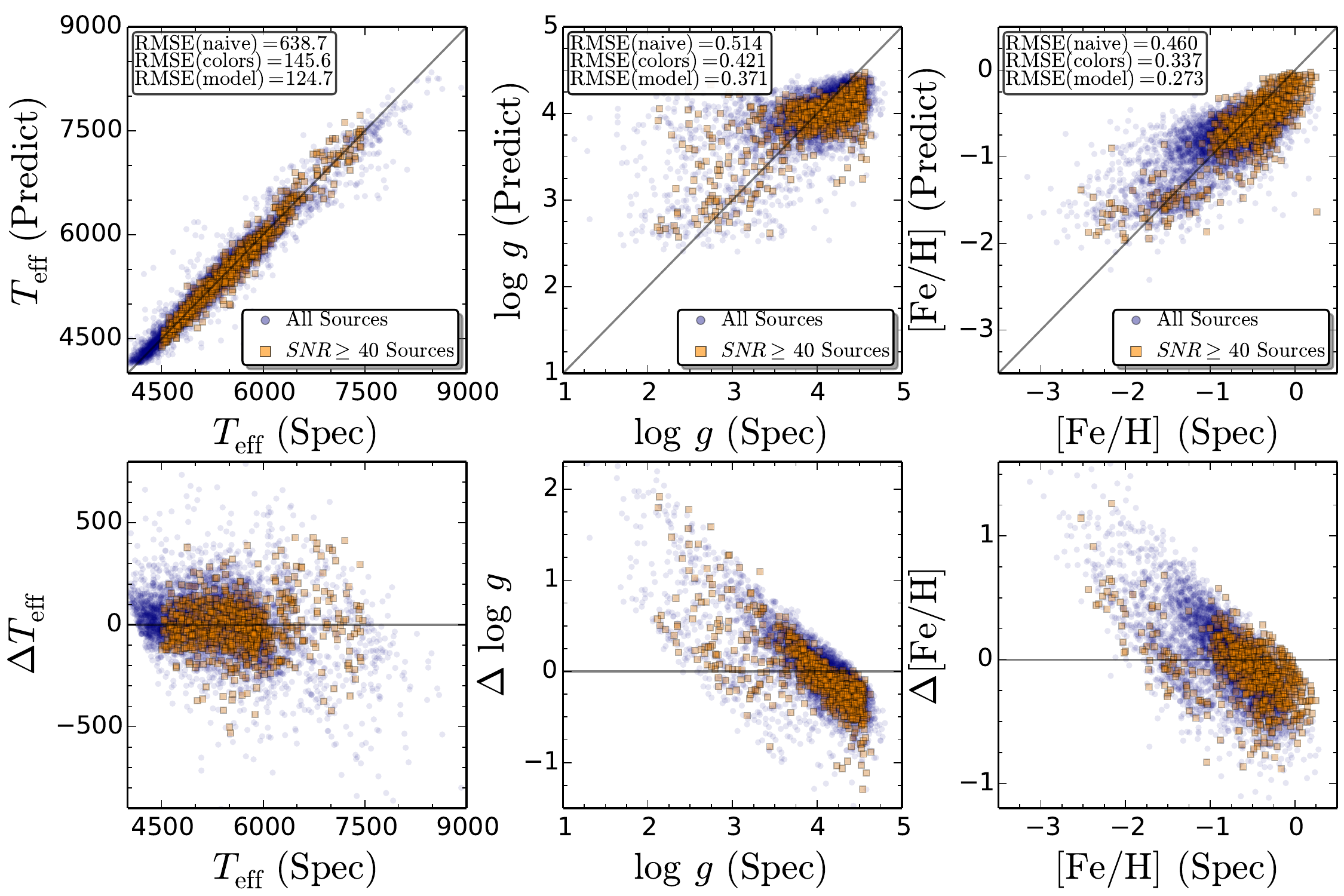}}
    \caption[Final random forest regression models for \teff, \logg, and \feh]{Random forest regression results for the final optimized models. \textit{Top}: Inferred values for all sources in the training set are shown as blue circles. The ``pristine sample'' identified in the text with 4500 K $< T_{\rm eff} <$ 7500 K, \logg\ $< 2$, and SNR $\geq 40$ (i.e.\ those with the most reliable SSPP parameter estimates) are shown as orange squares. The RMSE values quoted in the inset reflect the values for the pristine sample. For this subset of sources, the RMSE, a measure of the scatter in the model-estimated values of the parameters relative to their spectroscopic values, is $\sim$125 K, 0.37 dex, and 0.27 dex for \teff, \logg, and \feh, respectively. This performance is comparable to the typical uncertainties in these parameters associated with a low-resolution spectrum. Our final, optimized model significantly outperforms the naive model (where the predicted value for all sources is equal to the sample mean), and is $\approx$12--20\% better than an optimized random forest model using only single-epoch photometric colors. \textit{Bottom}: Residuals from the random forest regression model. There is a strong bias in the predictions for \logg\ and \teff. There are at least three sources contributing to this bias (one physical, two systematic); future improvements to the training set and model should reduce the bias in the final predictions (\S\ref{sec:bias}).}
    \label{fig:final_model}
\end{figure*}

We show the results of the final, optimized RFR models for predicting \teff, \logg, and \feh\ in Fig.~\ref{fig:final_model}. The cross-validated RMSE, a measure of the scatter between the predicted and true values of the parameters of interest, is 165 K, 0.39 dex, and 0.33 dex for \teff, \logg, and \feh, respectively. Our models show dramatic improvements of $\sim$78\%, 24\%, and 33\% for \teff, \logg, and \feh, respectively, over the naive model, where the predicted value for all sources is equal to the sample population mean. Our models further show an improvement of $\approx$7--9\% over optimized RFR models trained with single-epoch photometric colors (see Fig.~\ref{fig:model_compare}). 

The predictive power of the models improves significantly when examining subsets of the full training set. In addition to showing the cross-validated predictions for all sources in the training set, Fig.~\ref{fig:final_model} also highlights the predicted values for the subset of sources where the SSPP can be considered most reliable, namely, stars that did not raise any flags during either automated or visual inspection with 4500 K $< T_{\rm eff} <$ 7500 K, \logg\ $> 2$, and SNR $\geq 40$. The validation set of bright stars upon which SSPP was trained contained few examples outside this parameter space \citep{lee08}, which is why the predictions for hot and cool stars, as well as supergiants, are less reliable than those for main-sequence dwarfs. Furthermore, the predictive accuracy of the SSPP degrades rapidly as the SNR declines \citep{allende-prieto08}. For this ``pristine sample'' the RMSE is $\sim$125 K, 0.37 dex, and 0.27 dex for \teff, \logg, and \feh, respectively. \textit{This is comparable to the typical uncertainties associated with a low-resolution spectrum} \citep{lee08}, and it represents a $\approx$12--20\% improvement over colors-only models. Finally, we remind the reader that the cross-validated RMSE reflects the performance of the framework for sources that are similar to those in our training set. Thus, the application of this model to cool M stars or hot white dwarfs, stars for which the SSPP cannot provide reliable stellar parameters, will yield predictions that are significantly worse than the RMSE. 
Many, though not all, of the sources that fall outside our training set can be eliminated by adopting color cuts to select sources with 4500 K $<$ \teff\ $<$ 9000 K, the temperature range over which the SSPP is valid. 

While the focus of this work concerns the use of variability information to estimate stellar parameters, we note that for \teff\ the scatter in our colors-only model, $\sim$145 K, is superior to all but one \teff\ measurement method in the SSPP \citep{lee08}. This method is also vastly superior to the SSPP methods that rely only on color, which have typical scatter of $\sim$200 K. While future studies are certainly necessary, this suggests that efforts to automatically determine \teff\ (potentially for any stellar system, variable or otherwise) should consider our RFR model. 

Despite producing a scatter similar to estimates from low-resolution spectra, the regression models for \logg\ and \feh\ produce biased predictions, as can be seen in Fig.~\ref{fig:final_model}. There are at least three different factors contributing to this bias, which we discuss further in \S\ref{sec:bias}. An expanded training set containing more stars with low metallicity or low surface gravity, and having more precise spectroscopic determinations of the stellar parameters, will improve the future performance of model while also reducing the bias in the final predictions. Some physical effects (e.g., reddening, correlation between the parameters), however, may always prove difficult to overcome (\S\ref{sec:bias}). Additionally, the incorporation of machine learning regression tools capable of estimating uncertainties for their final predictions (e.g., \citealt{Wager13}) would help to identify sources with the most biased predictions.

\subsection{Are the Spectroscopic Samples from This Work and the SSPP Similar?}

The typical uncertainties for stellar parameters of bright stars (SNR $\ga$ 50) determined by the SSPP are $\sigma(T_{\rm eff}) = 157$ K, $\sigma(\log g) = 0.29$ dex, and $\sigma([{\rm Fe/H}]) = 0.24$ dex, over the temperature range 4500 K $\le$ \teff\ $\le$ 7500 K \citep{lee08}. While these uncertainties are similar to the RMSE values reported above for the pristine sample of our dataset, the methods used to determine the scatter as well as the underlying samples differ between this work and the SSPP. Here, we address those differences to determine how well our method compares to low-resolution spectroscopy. 

The final reported uncertainties for the SSPP are determined by adding internal and external uncertainties in quadrature. The external uncertainties dominate the error budget and are determined via a comparison of the SSPP parameters with parameters determined via high-resolution spectroscopy for a common sample of 125 stars \citep{lee08}. The scatter is determined via a gaussian fit to the residuals (e.g., \teff$_{\rm , \; SSPP}$~$-$~\teff$_{\rm , \; high-resolution}$). This method assumes that the tails of the distribution are gaussian, which is difficult to assess with only 125 stars. Thus, we prefer the RMSE, which makes no assumptions about the underlying distribution of uncertainties. In the case where the underlying distribution is gaussian, then the RMSE is approximately equal to the gaussian standard deviation. Using the parameters from the SSPP and the high-resolution-spectroscopic analysis, which are available in \citet{allende-prieto08}, we report the RMSE for the SSPP in Table~\ref{tab:sample-compare}. 

\begin{deluxetable}{lcccc}
\tabletypesize{\small}
\tablecolumns{5}
\tablewidth{0pt}
\setlength{\tabcolsep}{6pt}
\tablecaption{RMSE Comparison Between This Study and the SSPP Validation Set\label{tab:sample-compare}}
\tablehead{\colhead{} & \colhead{} &
    \colhead{This Work\tablenotemark{a}} & 
    \colhead{SSPP\tablenotemark{b}} & 
    \colhead{SSPP-boot\tablenotemark{c}} }
\startdata
    \teff\ (K)      & & 125 & 282\tablenotemark{d} & 289\tablenotemark{d} \\
    \logg\ (dex)    & & 0.37 & 0.35 & 0.31 \\
    \feh\ (dex)     & & 0.27 & 0.25 & 0.21 \\
\enddata
\vspace{-0.2cm}
\tablenotetext{a}{RMSE for the pristine sample of the dataset discussed in \S\ref{sec:opt_CV_results}.}
\tablenotetext{b}{RMSE comparing SSPP parameters to parameters measured from high-resolution spectra (see text).}
\tablenotetext{c}{RMSE for the weighted-bootstrap resamples of the SSPP data designed to approximate the distribution of stars in this study (see text).}
\tablenotetext{d}{For \teff, the RMSE for the SSPP reduces to $\sim$180 K once a systematic offset between the SSPP and high-resolution measurements is removed (see the text and \citealt{lee08} for further details).} 
\end{deluxetable}

For \logg\ and \feh, the RMSE for the pristine sample of our dataset and the SSPP are remarkably similar (see the middle two columns of Table~\ref{tab:sample-compare}). This comparison may be misleading, however, if the sample of stars in this study and those in the SSPP validation set are different. This scenario is likely given that the sources in this study are specifically selected from the population of variable stars. Thus, we employ a weighted bootstrap resampling method to better compare the stars from the SSPP validation set to those in this study. For both samples, we estimate the stellar population density individually in \teff, \logg, and \feh\ using a non-parametric gaussian kernel density estimator (KDE), where the KDE bandwidth has been determined via Scott's rule \citep{Scott92}. We then perform a weighted bootstrap resampling of the SSPP validation set, where the weights are determined via the ratio of the KDE estimate for our pristine sample to the KDE estimate of the SSPP validation set. The weights ensure that the bootstrap distribution of SSPP validation sources better matches the distribution of variable stars in our study. For each of the stellar parameters, we obtain 1000 bootstrap samples of the RMSE, and the mean of these RMSE values are reported in Table~\ref{tab:sample-compare}. Within the constraints of the currently available data, our weighted-bootstrap resampling produces a better comparison between our method and the SSPP. Ideally, future efforts to compare our method to parameters derived from low-resolution spectra would include a large sample of stars ($N \ga 1000$) which have high- and low-resolution spectroscopic observations in addition to well sampled light curves. 

Table~\ref{tab:sample-compare} shows that the RMSE for the SSPP samples is significantly worse than the method presented in this paper. The high-resolution analysis presented in \citet{allende-prieto08}, and adopted by \citet{lee08}, contains two sets of spectra, labeled ``HET,'' for those obtained with the Hoberly-Eberly Telescope, and ``OTHERS,'' for those obtained with other high-resolution instruments. The HET sample shows an approximately constant $\sim$$-$200 K bias relative to SSPP \teff\ measurements, while the OTHERS sample shows an approximately constant $\sim$40 K bias relative to the SSPP. When combined, these biases partially offset, leading to an overall scatter of $\sim$140 K, as adopted in \citet{lee08}. If we remove these approximately constant offsets from these respective samples, and reject 3-$\sigma$ outliers, we find that the RMSE for the SSPP measurements relative to the high-resolution analysis is $\sim$180 K. We echo the sentiment presented in \citet{lee08} that further high-resolution spectra, across the entire temperature range for the SSPP, should be obtained to further investigate this potential bias between high-resolution analyses and the SSPP. 

From Table~\ref{tab:sample-compare}, we see that the bootstrap sample of the SSPP validation set results in an RMSE scatter that is $\sim$18\% and 24\% better than the method in this paper for \logg\ and \feh, respectively. Thus, low-resolution spectroscopic observations are superior to our machine learning method. Nevertheless, our method, which produces a scatter that is only $\sim$20\% worse than low-resolution spectroscopy, remains competitive and very attractive as we embark upon the LSST era. Furthermore, we argue below (\S\ref{sec:bias}) that it should be possible to further refine and improve our machine-learning method. While broadband photometric methods will likely never serve as a replacement for low-resolution spectroscopy when it comes to measuring detailed atmospheric abundances (e.g., \citealt{lee11}), we have demonstrated that the combination of light curves and colors can produce reliable estimates of the fundamental atmospheric parameters \teff, \logg, and \feh.

\subsection{Final UWVSC predictions}\label{sec:final_preds}

As a final step in the construction of our model, we provide predictions of \teff, \logg, and \feh\ for each of the sources in the UWVSC. We exclude spectroscopically confirmed quasars from SDSS (see \citealt{Schneider10,Paris14}), and note that there may be additional quasars that have not yet been spectroscopically identified (see \citealt{butler11, MacLeod11a}). For the remaining sources, which will virtually all be stellar, our model predictions are provided for follow-up studies of Stripe 82 sources. We note that, similar to quasars, the predictions for sources outside the parameter space of our training set will likely be incorrect. The development of automated tools, similar to the SSPP, capable of measuring stellar parameters for hot (e.g., Wolf-Rayet stars), cold (e.g., pulsating red giants, flaring M-dwarfs), and very high surface gravity stars (e.g., white dwarfs), is essential for extending our model to cover the full range of variability types found across the Hertzsprung-Russell diagram. Our final predictions for UWVSC sources that are neither spectroscopically confirmed quasars nor sources with SSPP measured parameters (see \S\ref{sec:SSPP}) are provided in Table~\ref{tbl-final_preds}. Should any other researchers wish to design their own models using the training set from this study, we will happily make the data available upon request. 

\begin{deluxetable}{rrrrrr}
\tabletypesize{\small}
\tablecolumns{6}
\tablewidth{0pt}
\setlength{\tabcolsep}{6pt}
\tablecaption{Final Model Predictions for UW VSC Sources\label{tbl-final_preds}}
\tablehead{\colhead{UW ID\tablenotemark{a}} & \colhead{$\alpha_{\rm J2000.0}$} & 
    \colhead{$\delta_{\rm J2000.0}$} & \colhead{$T_{\rm eff}$} & \colhead{$\log \; g$} & \colhead{$[{\rm Fe/H}]$} \\
    \colhead{} & \colhead{} & \colhead{} & \colhead{(K)} & \colhead{(dex)} & \colhead{(dex)}}
\startdata
    429 &  1.145510 & -0.887604 &   4215 & 4.02 & -0.89 \\
   1606 &  3.407860 & -0.997584 &   4388 & 4.00 & -0.91 \\
   1970 &  0.599323 &  0.584808 &   5413 & 3.51 & -1.88 \\
   1990 &  4.362962 &  0.295144 &   4196 & 3.90 & -0.80 \\
   3219 &  0.826910 &  0.848085 &   5069 & 3.97 & -1.09 \\
   3271 &  0.653706 &  0.496230 &   6233 & 3.79 & -1.73 \\
   3352 &  0.347327 &  0.084452 &   7482 & 3.46 & -1.27 \\
   4058 &  1.636241 & -0.297831 &   4405 & 3.45 & -0.86 \\
   4357 &  0.054138 &  0.234712 &   4229 & 3.99 & -0.80 \\
   4400 &  0.592502 & -1.232759 &   4223 & 4.00 & -0.90 \\
\enddata
\vspace{-0.3cm}
\tablecomments{Only the first ten sources are presented here as an example of the form 
    and content of the complete table. The full table, containing all 53,781 sources with predicted stellar parameters 
    is available online.}
\tablenotetext{a}{Source ID in the UW VSC.}
\end{deluxetable}

\section{Discussion -- Understanding the Model Bias}\label{sec:bias}

There are three different effects that contribute to the bias present in the final predictions from the RFR models (see Fig.~\ref{fig:final_model}). The first is a consequence of using non-parametric, data-driven models: there is a natural regression to the mean wherein sources near the extremes of the population distributions are predicted to have values closer to the sample mean than their true values \citep{Zhang12}. The second is a biasing of the regression slope toward zero due to noise associated with the spectroscopically determined values of the stellar parameters, an effect known as regression dilution bias \citep{Frost00}. The third is a subtle physical effect associated with the correlation between color (i.e., \teff) and \logg\ and \feh\ in our training set sample. 

\subsection{Bias I: Regression to the Mean}

All non-parametric, data-driven regression methods experience regression to the training-set mean. For random forest regression, each decision tree predicts a parameter value for an unlabeled source that is equal to the mean value of that parameter for all training set sources that end up in the same terminal node of the tree as the unlabeled source. The final prediction for the parameter of interest is the mean of the values predicted by each of the individual trees. This methodology leads to two important consequences: the first is that random forest regression cannot make predictions outside the range of values contained within the training set, and the second is that sources located near the extrema of the training set will have predictions biased toward the training set mean. For instance, the star with the lowest surface gravity in our training set has $\log \, g ({\rm Spec}) = 0.705$ dex, but during cross validation the best possible prediction for \logg\ for this source is 1.39, the mean \logg\ value for the next six lowest surface gravity stars.\footnote{Recall that in the optimal \logg\ model $\texttt{nodesize} = 6$, meaning that every single tree prediction is the result of averaging the \logg\ values for \textit{at least} 6 training set sources.}

\begin{figure}
    \centerline{\includegraphics[width=3.6in]{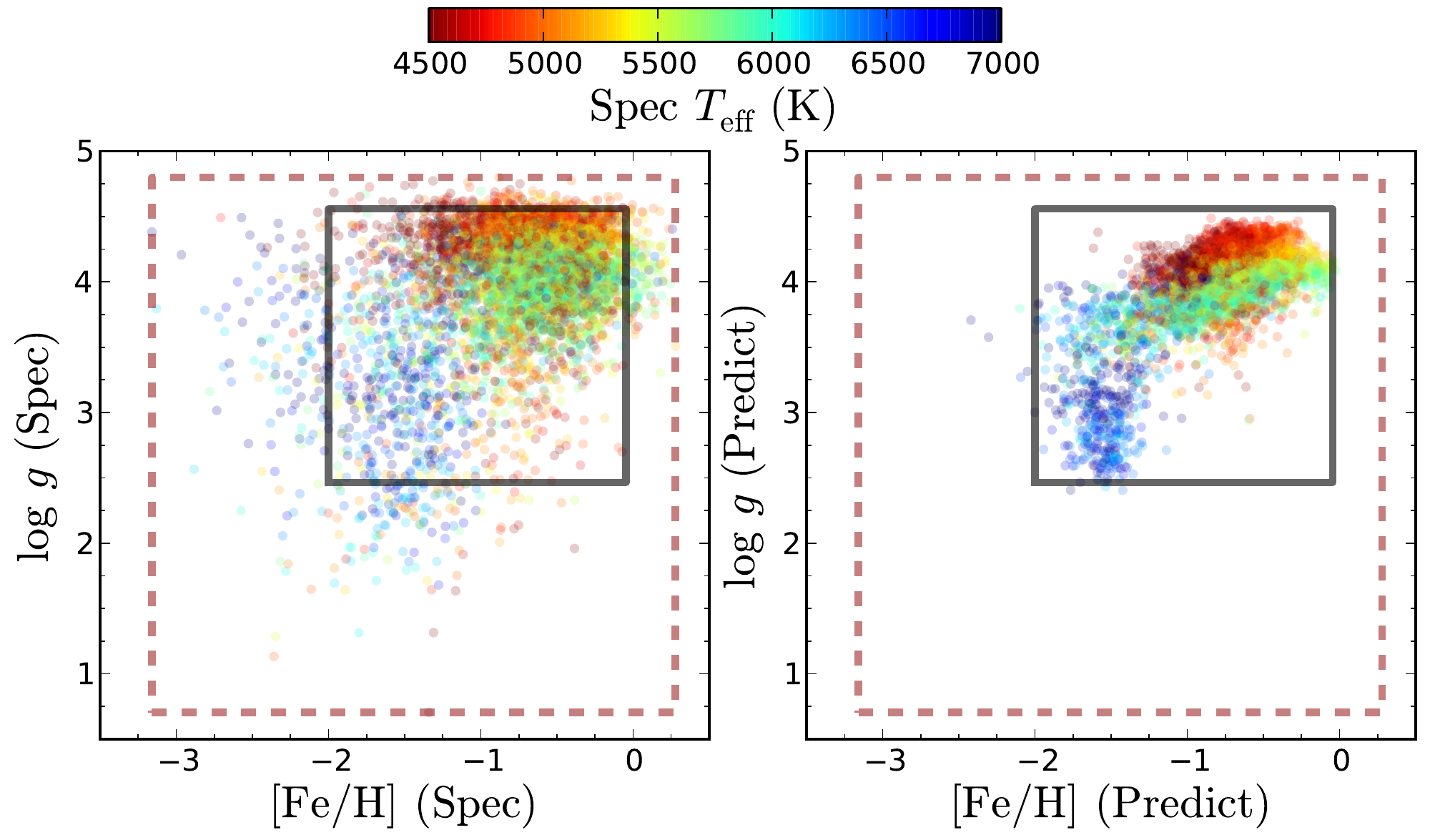}}
    \caption[Regression to the mean bias]{Distribution of all sources in our training set in the \logg--\feh\ plane, showing their spectroscopically measured values (\textit{left}), and their random forest regression-inferred values (\textit{right}). The points have been colorized according to their spectroscopically-measured \teff. The dashed red line shows the full ranges of values represented in the training set, while the solid grey line shows the inter $95^{\rm th}$ percentile of the training set distributions. Notice that virtually all of the model-inferred predictions lie inside the inter $95^{\rm th}$ percentile, which is strongly biased towards high \logg\ and high \feh. This regression to the mean results in predictions for sources near the low (high) end of the distribution to be over- (under-) predicted, which biases the regression slope towards zero.}
    \label{fig:reg_to_mean}
\end{figure}

To further illustrate the results of this bias, a scatter plot showing both the spectroscopic and model-predicted values of \logg\ against \feh\ is shown in Fig.~\ref{fig:reg_to_mean}. The red dashed lines show the extrema of the training set distributions: the random forest regression model is incapable of predicting values outside this area. Even more revealing is the inter $95^{\rm th}$ percentile of the training set distributions, shown via solid grey lines. The right side of Fig.~\ref{fig:reg_to_mean} shows that virtually all random forest regression predictions are encompassed by the inter 95\% of the training set distribution. Thus, sources with low (high) values of \logg\ or \feh\ will have predictions that are biased above (below) the true values of those parameters.

One way to improve this bias would be to reduce the imbalance that currently exists within the training set. The scarcity of sources with low surface gravity or low metallicity makes it difficult for the model to predict similar values for unlabeled sources. A reduction in the imbalance in the training set would effectively expand the size of the grey rectangle shown in Fig.~\ref{fig:reg_to_mean}, lowering the bias in predictions for sources near the extrema of the training set distribution. It is possible that this reduction in the bias could lead to increased variance in the model predictions, potentially increasing the RMSE. In an ideal scenario, the training set sample would be fully representative of the population of sources for which predictions are required, but this is often difficult to achieve in practice. 

\subsection{Bias II: Regression Dilution}

The second factor contributing to the final-model bias is the result of both the light-curve features and the spectroscopically-measured parameter values being noisy. The uncertainties associated with these measurements leads to a flattening of the regression slope, an effect known as regression dilution bias (e.g. \citealt{Frost00}). Moving forward, it will be impossible to completely eradicate this bias since infinitely precise light-curve feature measurements and spectroscopic parameter measurements will never be available. Nevertheless, there are some improvements that could be made to mitigate against this bias. 

Light curve features, particularly those that measure periodicity - the most important features for variable star classification \citep{richards11,Dubath11} - are strongly dependent upon both the observational cadence and photometric noise properties of a survey. There is empirical evidence that the uncertainty on the light-curve feature measurements decreases if the total number of observations increases or the photometric accuracy improves \citep{Graham13}. With the caveat that the samples are relatively small, this also appears to be true in the present study: for \teff\ and \feh\ we find that the RMSE for the subsample of pristine sources with more than 40 $g$-band observations is $\sim$15\% better than for pristine sources with fewer than 30 $g$-band observations. The improvement is less dramatic in \logg\ ($\sim$5\%), but we find a $\sim$20\% improvement for sources with clearly identified periodicity. This portends well for LSST, which will obtain $\sim$1000 observations per star, significantly more than the SDSS light curves in the present study, which have a median of 33 $g$-band observations. 

More precise spectroscopic measurements can be obtained, either via high-resolution spectroscopy or an automated method that improves upon the SSPP. However, as the nature of our sample is selected due to its variability, the true values of \logg\ or \teff\ cannot be precisely measured in a single instant (i.e.\ the time of spectroscopic observation), because they are variable. One possible, though expensive, way to reduce this bias is via repeated observations of a subset of stars in the training sample, which can then be used to estimate a multiplicative factor to correct the regression dilution bias \citep{Frost00}.

\subsection{Bias III: Correlation of \logg\ and \feh\ with Photometric Color}

The third, and possibly most significant, contribution to the observed bias in the final predictions for \logg\ and \feh\ is the importance of the color features. Figures~\ref{fig:logg_ff}~and~\ref{fig:feh_ff} show that the four color features are among the most powerful discriminants for predicting \logg\ and \feh, while Fig.~\ref{fig:loggVteff} shows that the cooler stars ($T_{\rm eff} \le 6000$ K) in our training set are most likely to have a dwarf-like surface gravity ($\log \, g \approx 4.2$) and sub-solar metallicity (\feh\ $ \approx -0.6$), while warm stars ($T_{\rm eff} \geq 6500$ K) are most likely to have giant-like gravity ($\log \, g \approx 3.5$) and be metal poor (\feh\ $ \approx -1.3$).

\begin{figure*}[t]
    \centerline{\includegraphics[width=6in]{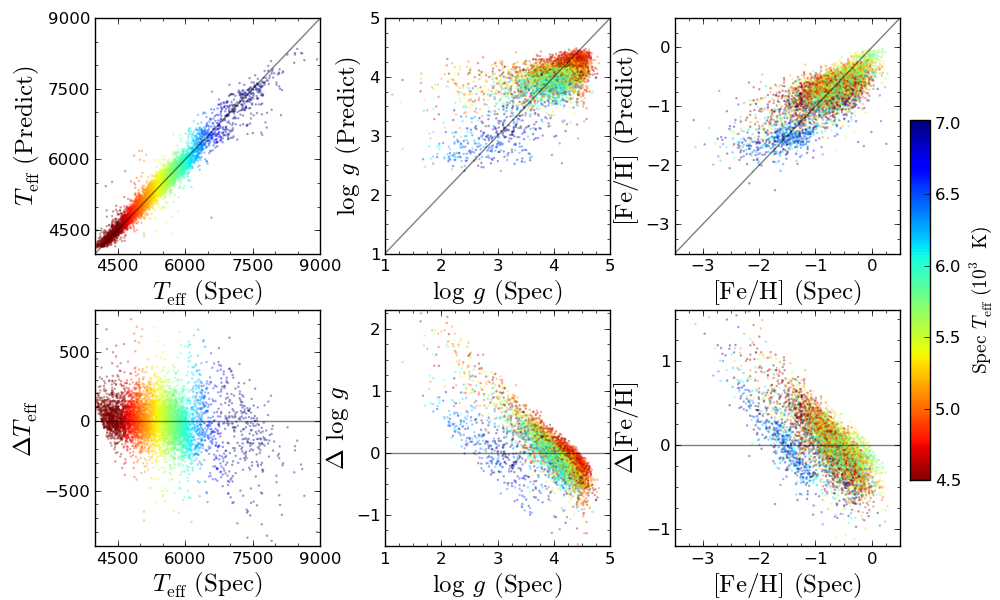}}
    \caption[Bias due to photometric colors]{Recreation of Fig.~\ref{fig:final_model} showing \textit{all} sources considered in this study colored according to their spectroscopically-measured \teff\ (a proxy for photometric color). For \logg\ and \feh\ there is a clear temperature-dependent substructure in the final predictions. This temperature dependance leads to biased final predictions (see text).}
    \label{fig:color_bias}
\end{figure*}

To further illustrate the importance of photometric colors, we once again show our final model predictions in Figure~\ref{fig:color_bias}, with the individual sources colored via their spectroscopically determined \teff. It is immediately clear from this figure that almost without exception warm stars are predicted to have low \logg\ and low \feh, while the opposite is true for cooler stars. The photometric colors essentially restrict the random forest regression predictions of \logg\ and \feh\ to a narrow range of values, which can only be slightly refined by the light curve features. In other words, once the model recognizes a source as cool, it is incapable of then identifying it as either a low surface gravity or low-metallicity star.

To test this hypothesis we constructed a random forest regression model to predict \logg\ on the subset of sources within our training set with $g - r > 0.6$, roughly corresponding to $T_{\rm eff} < 5000$ K. The model was constructed without photometric colors, i.e.\ using only the time-domain features. The performance of this model is worse ($\approx 20$\%) than the final optimized model presented in \S\ref{sec:final_results}. The same is true for weighted random forest regression models. We conclude that the 66 light-curve features adopted for this study are incapable of clearly distinguishing between low surface gravity sources and dwarf-like surface gravity sources at similar \teff. Fig.~\ref{fig:color_bias} shows that the \feh\ predictions are affected by a similar phenomenon, though the effect is less pronounced in that case. 

While the light-curve features are incapable of clearly separating supergiants from dwarfs at similar temperatures, it is important to remember that the addition of light-curve features does improve the overall performance of the models, by as much as $\approx$20\%. Moving forward, the best way to alleviate the photometric-color bias would be via the creation of new features capable of separating high and low \logg\ (\feh) sources at similar temperatures, though we note that feature construction is a very difficult problem. We believe the optimal approach to improve the feature set would be to perform a full classification of the variables in Stripe 82, similar to the classification of sources in the All Sky Automated Survey \citep{Richards12a}. Following classification, domain experts on the class of stars comprising the low \logg\ (\feh) and high \logg\ (\feh) stars could help to identify new features capable of separating these variables at similar temperatures. Adding these features to the regression models should significantly improve its overall predictive power.

\section{Conclusions}

We have presented a new machine-learning framework that is capable of predicting \teff, \logg, and \feh\ from photometric observations alone. The framework was built following a systematic spectroscopic survey of variable sources. Targets for the spectroscopic survey were selected from the UWVSC, a publicly available catalog of SDSS light curves for \stripe\ variable sources. The survey was designed to be agnostic toward variability class: all bright ($r \le 19$ mag), well-observed ($\ge 24$ epochs) sources were included as potential targets. 

Spectroscopic observations were carried out using the multi-fiber Hectospec instrument on the MMT. In sum, we obtained 5914 Hectospec spectra of 5825 unique sources, which we supplemented with 3121 SDSS spectra of stellar sources in the UWVSC. We applied an adapted version of the SSPP to each of the 9035 spectra in our combined sample to determine \teff, \logg, and \feh. The SSPP produced reliable estimates of these parameters for 5994 sources; the remaining sources suffered some peculiarity (most often low SNR or \teff$ < 4000$ K) that prevented reliable estimates of their stellar parameters. 

To characterize the photometric behavior of UWVSC sources, we measured 66 light-curve features in each of the $ugriz$ bands. Our machine-learning framework utilizes the random forest algorithm to perform a non-parametric regression between \teff, \logg, and \feh\ and these photometric light-curve features. Thus, we have developed a method capable of measuring these parameters without the need for spectroscopic observations. Our final, optimized models determine \teff, \logg, and \feh\ with an RMSE of 165 K, 0.39 dex, and 0.33 dex, respectively. When we restrict our sample to the subset of sources for which the SSPP is most reliable the RMSE decreases to $\sim$125~K, 0.37~dex, and 0.27~dex, respectively. This scatter is comparable to what is achieved with low-resolution spectra, and it represents an improvement of $\approx$12--20\% over machine learning models trained solely with photometric colors. The model predictions of \logg\ and \feh\ are biased as a result of three different effects: (i) regression to the mean, (ii) regression dilution bias, and (iii) a physical effect associated with the correlation between photometric colors and \logg\ and \feh\ in our training set. We discussed possible methods to alleviate these biases in the future. 

We view the results presented herein as an important step towards the goal of extracting the most impactful information from photometric time-domain surveys. The UWVSC contains $\sim$67k sources, in contrast, LSST is expected to discover at least 50 million variable stars \citep{Ivezic08}. The vast majority of these sources will be prohibitively faint for spectroscopic observations on anything smaller than a 30-m class telescope. This necessitates the development of novel software applications: as a demonstration of our framework, we presented estimates of \teff, \logg, and \feh\ for all UWVSC sources (see  \S\ref{sec:final_preds}).

As we embark upon the burgeoning age of celestial cinematography (the LSST will, in essence, make a 10 yr movie, sampled every 3 days, of everything in the southern sky), it is essential that we develop advanced tools for discovery. More data does not always equate to better information nor expanded knowledge: sophisticated new tools are required to decipher the complex data stream from LSST. Based on the results shown here, it is not unreasonable to think that LSST may be considered a pseudo-spectrographic engine. Our machine-learning framework will allow fundamental parameters to be determined without the need for additional spectroscopy. The method can be leveraged for a huge advantage given the high data rates of upcoming surveys and the difficulty involved for spectroscopic follow-up. Once the atmospheric parameters are determined additional fundamental properties, namely mass $M_\ast$, luminosity $L_\ast$, and radius $R_\ast$, can be (probabilistically) inferred (e.g., \citealt{Schoenrich13}). In this way the most detailed maps of the Milky Way ever constructed will be charted, which promises to reveal and solve several mysteries regarding the formation of the Galaxy.   

\acknowledgments 

This work has made extensive use of the online data and tools 
made available by the SDSS collaboration. We are particularly 
grateful to {\v Z}.~{Ivezi{\'c}} and collaborators at the University of 
Washington for making their calibrated light curves of Stripe 82 sources 
publicly available. We thank BD Bue for a fruitful conversation 
concerning regression bias. We also thank the anonymous referee for several 
useful comments that have helped to improve this paper. 

A.A.M. acknowledges support for this work by NASA from a
Hubble Fellowship grant: HST-HF-51325.01, awarded by STScI,
operated by AURA, Inc., for NASA, under contract NAS 5-26555. 
J.S.B. acknowledges support from an NSF-CDI grant 0941742.
JAE gratefully acknowledges support from an Alfred P. Sloan Research Fellowship. 
Part of the research was carried out at the Jet Propulsion 
Laboratory, California Institute of Technology, under a contract
with NASA. 
Observations reported here were obtained at the MMT Observatory, a joint 
facility of the University of Arizona and the Smithsonian Institution.

{\it Facilities:} 
\facility{Sloan}, \facility{MMT (Hectospec)}

\end{document}